\documentclass{aa}
\usepackage{graphicx}
\usepackage{txfonts}
\usepackage{natbib}
\usepackage[breaklinks=true,colorlinks=true,linkcolor=blue,citecolor=blue]{hyperref}
\usepackage[squaren,Gray]{SIunits}
\newcommand{\mjup}{\ensuremath{\textnormal{M}_{\textnormal{Jup}}}}
\newcommand{\msun}{\ensuremath{\textnormal{M}_{\odot}}}
\newcommand{\rd}{\mathrm{d}}
\newcommand{\excs}{\extracolsep{\fill}}
\newcommand{\bp}{$\beta\:$Pic}
\newcommand{\bpb}{$\beta\:$Pic~b}
\newcommand{\bpc}{$\beta\:$Pic~c}
\newcommand{\bpw}{$\beta\:$Pictoris}
\newcommand{\micron}{\unit{}{\micro\meter}}
\begin{document}
\title{Dynamics of the \bpw\ planetary system and its falling evaporating bodies}
\author{H. Beust\inst{1} \and  J. Milli\inst{1} \and A. Morbidelli\inst{3} \and S. Lacour\inst{2,3} \and A.-M. Lagrange\inst{2} \and G. Chauvin\inst{4} \and M. Bonnefoy\inst{1} \and J. Wang\inst{5}}
\institute{$^{1}$Univ. Grenoble Alpes, CNRS, IPAG, 38000 Grenoble, France\\
 $^{2}$LESIA, Observatoire de Paris, Universit\'e PSL, CNRS, Sorbonne Universit\'e, Université de Paris Cit\'e, 5 place Jules Janssen, 92195 Meudon, France\\
$^{3}$   European Southern Observatory, Karl-Schwarzschild-Straße 2, 85748 Garching, Germany \\
$^{4}$Universit\'e C\^ote d’Azur, Observatoire de la C\^ote d’Azur, CNRS, Laboratoire Lagrange, France\\
$^{5}$Center for Interdisciplinary Exploration and Research in Astrophysics (CIERA) and Department of Physics and Astronomy, Northwestern University, Evanston, IL 60208, USA
}
\date{Received ....; Accepted....}  \offprints{H. Beust}
\mail{Herve.Beust@univ-grenoble-alpes.fr}
\titlerunning{Dynamics of the \bpw\ planetary system}
\authorrunning{H. Beust et al.}
\abstract{
For decades, the spectral variations of \bpw\ have been  modelled as the result of the evaporation of 
exocomets close to the star, termed falling evaporating bodies (FEBs). Resonant perturbations by a
hypothetical giant planet have been proposed to explain the dynamical origin of these stargrazers. The disk is
now known to harbour two giant planets, \bpb\ and c, orbiting the star at 9.9\,au and 2.7\,au. While the former almost matches the planet formerly suspected, the recent discovery of the latter complicates the picture.}
{We first question the stability of the two-planet system. Then we investigate the dynamics of a disk of
planetesimals orbiting the star together with both planets to check the validity of the FEB generation mechanism.}{Symplectic N-body simulations are used to first determine which regions of the planetesimal disk are dynamically stable and which are not. Then we focus on regions where disk particles are able to reach high eccentricities, mainly thanks to resonant mechanisms.}
{The first result is that the system is dynamically stable. Both planets may temporarily fall in 7:1 mean motion 
resonance (MMR). Then, simulations with a disk of particles reveal that the whole region extending between
$\sim1.5\,$au and $\sim 25\,$au is unstable to planetary perturbations. However, a disk below 1.5\,au survives, which
appears to constitute an active source of FEBs via  high-order MMRs with \bpc. In this new picture, \bpb\ acts as a distant perturber that helps sustain the whole process.}
{Our new simulations rule out the preceding FEB generation mechanism model,  which placed their origin
at around 4--5\,au. Conversely, FEBs are likely to originate from a region much further in and related to MMRs with \bpc. That mechanism also appears to last longer, as new planetesimals are able to continuously enter the MMRs and evolve towards the  FEB state. Subsequently, the physical nature of the FEBs may differ from that  previously thought, and presumably may not be icy.}
\maketitle
\section{Introduction}
The southern A6V \citep{2006AJ....132..161G} star \bpw\ (\bp) has been the focus of intense research for decades, initially for its wide edge-on debris disk \citep{1984Sci...226.1421S}. The exact age of this young main-sequence star has been the subject of controversy. The most robust determinations \citep[$22\pm6\,$Myr][]{2014MNRAS.438L..11B,2017AJ....154...69S} \citep[$18.5^{+2.0}_{'-2.4}\,$Myr][]{2020A&A...642A.179M} are based on the young moving group it belongs to. In the following we assume an age of 20\,Myr as a typical standard value.

The presence of hidden planets in the circumstellar disk was suspected very early, as a natural outcome of the disk evolution. Specific clues suggesting the presence of at least one giant planet were identified. First, the warped profile of the disk as well as its various asymmetries
\citep{1995AJ....110..794K,2000ApJ...539..435H} were tentatively attributed to the perturbing action of a planet \citep{1997MNRAS.292..896M,2001A&A...370..447A}. Similarly, the photometric variations observed in 1981
\citep{1995A&A...299..557L} were also tentatively attributed to a planetary transit \citep{1997A&A...328..311L}. Finally, the repeated transient spectral variations monitored in the \bp\ absorption spectrum, interpreted as resulting from the evaporation of star-grazing exocomets \citep{1996A&A...310..181B} (see details below), were also attributed to the perturbing action of a giant planet \citep{1996Icar..120..358B,2000Icar..143..170B,2001A&A...376..621T}. These independent studies all agreed on a giant planet orbiting \bp\ at $\sim 10\,$au.

This planet, today known as \bpb, was indeed detected in high-contrast imaging first with NaCo \citep{2010Sci...329...57L}, and later with GPI \citep{2016AJ....152...97W} and SPHERE \citep{2019A&A...621L...8L}. It was thus immediately tempting to identify \bpb\ with the previously suspected planet. Before being able to give a robust conclusion, an accurate knowledge of \bpb's orbit was necessary. This was done thanks to a regular astrometric follow-up of the planet \citep{2012A&A...542A..41C,2014A&A...567L...9B,2016AJ....152...97W,2019A&A...621L...8L}. Currently, these studies all converge towards a semi-major axis distribution peaking close to 9--10\,au and an orbital period of $\sim20$\,yr. The eccentricity is harder to constrain, but it is presumably small.

The mass of \bpb\ is less easy to constrain. Fitting the relative astrometric data, it is only possible to fit the total dynamical mass of the system, which falls in the range $1.75$--$1.85\msun$ \citep{2016AJ....152...97W,2019ApJ...871L...4D}. This is in fact no more than a fit of the central star's mass. Photometric constraints deduced from direct imaging and models, combined with radial velocity data, helped to derive an estimate around $~10\,\mjup$. Based on a joint analysis of \textsl{Gaia} and \textsl{Hipparcos} data, \citet{2018NatAs...2..883S} report a mass of $11\pm2\,\mjup$, while \citet{2019ApJ...871L...4D} derive $13.1^{+2.8}_{-3.2}\,\mjup$.

All these characteristics were closely compatible with those deduced from the former theoretical studies listed above for the hypothetical planet orbiting \bp, so that the whole picture appeared clear. Meanwhile, a careful analysis of the HARPS radial velocity of the star led to the identification of another massive planet  termed \bpc\ orbiting the star at 2.7\,au \citep{2019NatAs...3.1135L} (i.e. well inside \bpb's orbit). Thanks to high-precision astrometric data from the GRAVITY interferometer, that planet was  directly confirmed \citep{2020A&A...642L...2N}, so that its reality is now certain. All orbital determinations aim now at fitting the two planetary orbits together \citep{2020A&A...642A..18L,2020A&A...642L...2N,2021A&A...654L...2L}, making use of available astrometric data for both planets and radial velocity data of the star. Currently, \bpc\ appears slightly less massive ($\sim8\,\mjup$) 
than \bpb, but significantly more eccentric, with an eccentricity around 0.3.

Gas in the circumstellar disk of \bp\ was detected very early in the stellar spectrum \citep{1985ApJ...293L..29H}, thanks to the edge-on orientation of the disk. Repeated observations of various metallic species \citep[\ion{Ca}{ii}, \ion{Mg}{ii}, \ion{Al}{iii}][]{1994A&A...290..245V} rapidly revealed frequent and transient changes appearing as additional absorption components in the bottom of the lines, most of the time redshifted by tens to hundreds of km/s with respect to the stellar velocity, and with variation  timescales of a few days or even less \citep{1987A&A...185..267F,1992A&A...264..637L,1999MNRAS.304..733P,2018MNRAS.479.1997K,2019MNRAS.489..574T}. These events have been convincingly modelled as the result of the sublimation of star-grazing exocomets in the vicinity of the star, each transient event corresponding to the passage across the line of sight of a single body \citep{1990A&A...236..202B,1996A&A...310..181B,1998MNRAS.294L..31C}. This scenario has been termed falling evaporating bodies (FEBs), and it implies an activity of hundreds of such FEBs per year having star-grazing periastron passages. These bodies do not actually graze the stellar surface at each periastron passage. They only need to be close enough to the star to allow their refractory compounds to sublimate and produce the observed absorption components. As shown by \citet{1998A&A...338.1015B}, in the environment of an A-type star like \bp, this occurs below a threshold distance of $\sim0.5\,$au (i.e. far above the stellar surface). For comparison purposes, this threshold distance around the Sun is no more than a few solar radii. Any exocomet orbiting \bp\ that gets closer to the central star than this distance is susceptible to becoming a FEB. The various subsets or families of variable components (in terms of velocity, depth, and variation timescales) have been successfully interpreted as FEBs crossing the line of sight at different distances within the quoted threshold, down to a few stellar radii only \citep{1996A&A...310..181B,2014Natur.514..462K}. The reality of this scenario was recently reinforced by the detection of similar events in transit photometry with the TESS satellite \citep{2019A&A...625L..13Z,2022NatSR..12.5855L}. A detailed analysis of the size distribution of the exocomets even revealed a differential distribution close to that of the collisional equilibrium. 

In this framework, the redshift (or blueshift) velocity of each transient component corresponds to the projection of the FEB's orbital velocity onto the line of sight at the time it crosses it, as what we observe is basically a cloud of metallic ions surrounding it. If we assume that these bodies move on very elongated orbits that can be assimilated to parabolic around periastron passage, this velocity mainly depends on the periastron distance of the FEBs and on the longitude of their periastron with respect to the line of sight. For instance, any redshifted (resp. blueshifted) component corresponds to a FEB crossing the line of sight before (resp. after) periastron. The large number of FEB events recorded \citep[see e.g.][]{2019MNRAS.489..574T} subsequently allows one to draw statistics on the FEB population \citep{2014Natur.514..462K}. The most obvious outcome is the statistical difference between redshifts and blueshifts. Although a few blueshifted events have been reported, the variable events appear most of the time on the red wing of the spectral lines. This shows up in all observational campaigns quoted above. This indicated a non-uniform distribution of the longitudes of periastra. The FEBs actually seem to come from one specific side of the disk \citep{1990A&A...236..202B}. Further detailed studies were able to identify several subfamilies with different orbital characteristics \citep{1998A&A...338.1015B,2014Natur.514..462K}, but this main result still holds. 

These constraints challenge any dynamical model willing to explain this high number of star-grazers. The common feature of all models is that planetary perturbations are needed to trigger the FEB phenomenon. For instance the Kozai-Lidov resonance, often presented as a source of star-grazers in the Solar System \citep{1992A&A...257..315B}, cannot be invoked here because of its natural rotational invariance. In this context, there should be as numerous redshifted as blueshifted events. Conversely, mean-motion resonances (MMRs) represent a potential anisotropic source of stargrazers. This idea was first investigated by \citet{1989A&A...213..436Y}, who showed that fictitious bodies trapped in some inner MMRs such as 4:1, 3:1, 5:2, 7:3, and 7:4 with a moderately eccentric giant planet can undergo large resonant eccentric increases. This idea was further developed by \citet{1996Icar..120..358B,2000Icar..143..170B} and \citet{2001A&A...376..621T} in the context of the FEB model for \bp. Semi-analytical and numerical studies showed that this model was indeed able to efficiently trigger the FEB phenomenon and reproduce nearly all their observed dynamical characteristics, provided the planet assumes an eccentricity $e\ga0.05$, and has a suitable value $\varpi=-70\degr\pm20\degr$ of longitude of periastron with respect to the line of sight. Detailed modelling \citep{2000Icar..143..170B,2001A&A...376..621T} led to the conclusion that the planet was presumably orbiting the star at $\sim 10\,$au within 50\%\ uncertainty. The mass of the perturbing planet was less easy to constrain, as the topology of the secular dynamics involved is nearly independent of it. Recently, this dynamical mechanism was theoretically re-investigated by \citet{2017A&A...605A..23P} who confirmed its general efficiency, showing that it is still active even if the perturbing planet is significantly more eccentric.

This resonant mechanism could apply in other systems than \bp. The Kirkwood gaps in the  solar asteroid belt have been emptied by a similar process in its early evolution \citep{1993Icar..102..316M,1995Icar..114...33M,1994Natur.371..314F}. The \bp\ system presumably represents a likely analogue to the young Solar System.

As mentioned above, the first detected planet \bpb\ closely matches the characteristics of the putative planet of 2001. However, the presence of the second planet \bpc\ orbiting inside complicates the picture. Any FEB progenitor initially trapped in an inner MMR with \bpb\footnote{4:1 or 3:1, located respectively at $\sim 4\,$au and $\sim 5\,$au from the star} orbits initially outside \bpc's orbit. During the resonant eccentricity increase process that drives it to the FEB state, it inevitably comes to regularly cross that orbit as the semi-major axis only undergoes small changes \citep{2017A&A...605A..23P}. Subsequent close encounters with \bpc\ could presumably eject it well before reaching the FEB state. \citet{2000Icar..143..170B} already investigated the possible perturbing action on FEB progenitors of a second planet orbiting inside the first one, but that planet was supposed to be Earth-sized. Such a small planet was indeed able to divert some of the FEB progenitors from their resonant dynamical route, but most of them were still able to safely reach the FEB state. Here \bpc\ is much more massive and could actually kill the resonant mechanism.

In the present paper we re-investigate the resonant FEB scenario based on the present day knowledge of the dynamical configuration of the \bp\ planetary system. In Sect.~2 we first investigate the stability and dynamical perturbations of the two-planet system, taking as input basis the latest orbital determination by \citet{2021A&A...654L...2L}. Then in Sect.~3 we add a disk of test particles to the simulation to see where FEB progenitors can be stable. Then we focus on the surviving part of the disk inside \bpc's orbit, showing that it can be a valuable source of FEBs in the presence of both planets, leading to a revision of the scenario. In Sect.~4 we discuss the consequences of the revised model. Our conclusions are presented in Sect.~5.
\section{Two-planet secular dynamics}
\begin{table}
\caption[]{Orbital parameters (Jacobi coordinates) of the two solutions integrated, following the conventions of \citet{2021A&A...654L...2L} and \citet{2020AJ....159...89B}. The reference epoch for the initial orbital phase $\tau$ is MJD 2,459,000 (May 21, 2020).}
\label{orbits}
\begin{tabular*}{\columnwidth}{@{\excs}lll}
\hline\noalign{\smallskip}
Parameter & Solution \#1 & Solution \#2\\
\noalign{\smallskip}\hline\noalign{\smallskip}
Stellar mass $M_*$ ($\msun$) & 1.77 & 1.85\\
\noalign{\smallskip}\hline\noalign{\smallskip}
\bpc~:
Mass $M_c$ ($\mjup$) & 8.41 & 8.85 \\
Semi-major axis $a_c$ (au) & 2.70 & 2.60 \\
Orbital period $P_c$ (yr) & 3.327 & 3.162 \\
Eccentricity $e_c$ & 0.33 & 0.33 \\
Inclination $i_c$ (deg) & 88.95 & 88.82 \\
Long. of asc. node $\Omega_c$ (deg) & 31.05 & 31.06 \\
Arg. of periastron $\omega_c$ (deg) & 67.70 & 61.02 \\
Orbital phase $\tau_c$ & 0.73 & 0.71 \\
\noalign{\smallskip}\hline\noalign{\smallskip}
\bpb~:
Mass $M_b$ ($\mjup$) & 11.73 & 10.00 \\
Semi-major axis $a_b$ (au) & 9.91 & 9.95 \\
Orbital period $P_b$ (yr) & 23.323 & 23.604 \\ 
Eccentricity $e_b$ & 0.10 & 0.10 \\
Inclination $i_b$ (deg) & 88.99 & 88.98 \\
Long. of asc. node $\Omega_b$ (deg) & 31.81 & 31.81 \\
Arg. of periastron $\omega_b$ (deg) & 196.10 & 202.05 \\
Orbital phase $\tau_b$ & 0.71 & 0.73 \\
\noalign{\smallskip}\hline
\end{tabular*}
\end{table}
\begin{figure*}
\makebox[\textwidth]{
\includegraphics[width=0.33\textwidth]{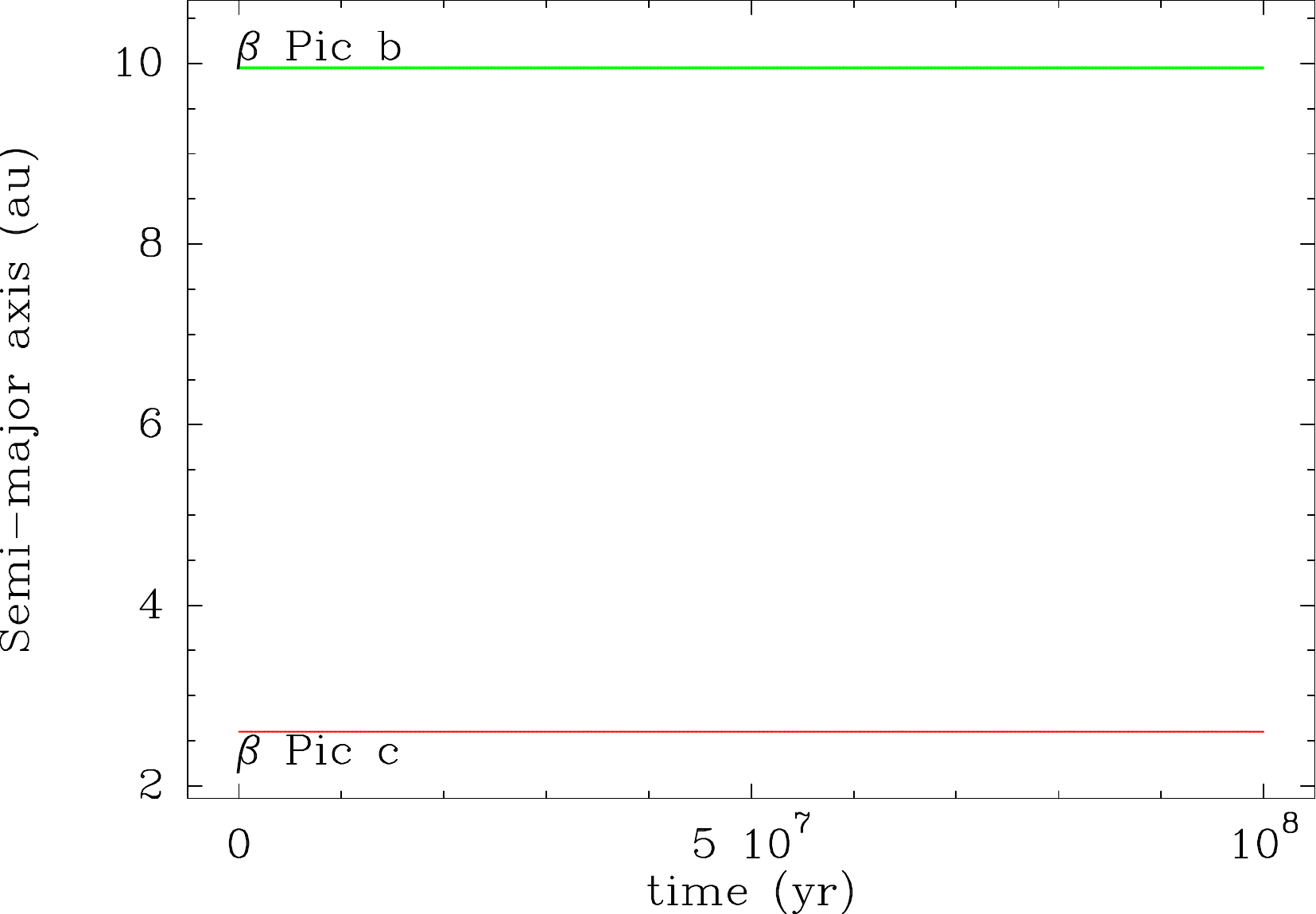} \hfil
\includegraphics[width=0.33\textwidth]{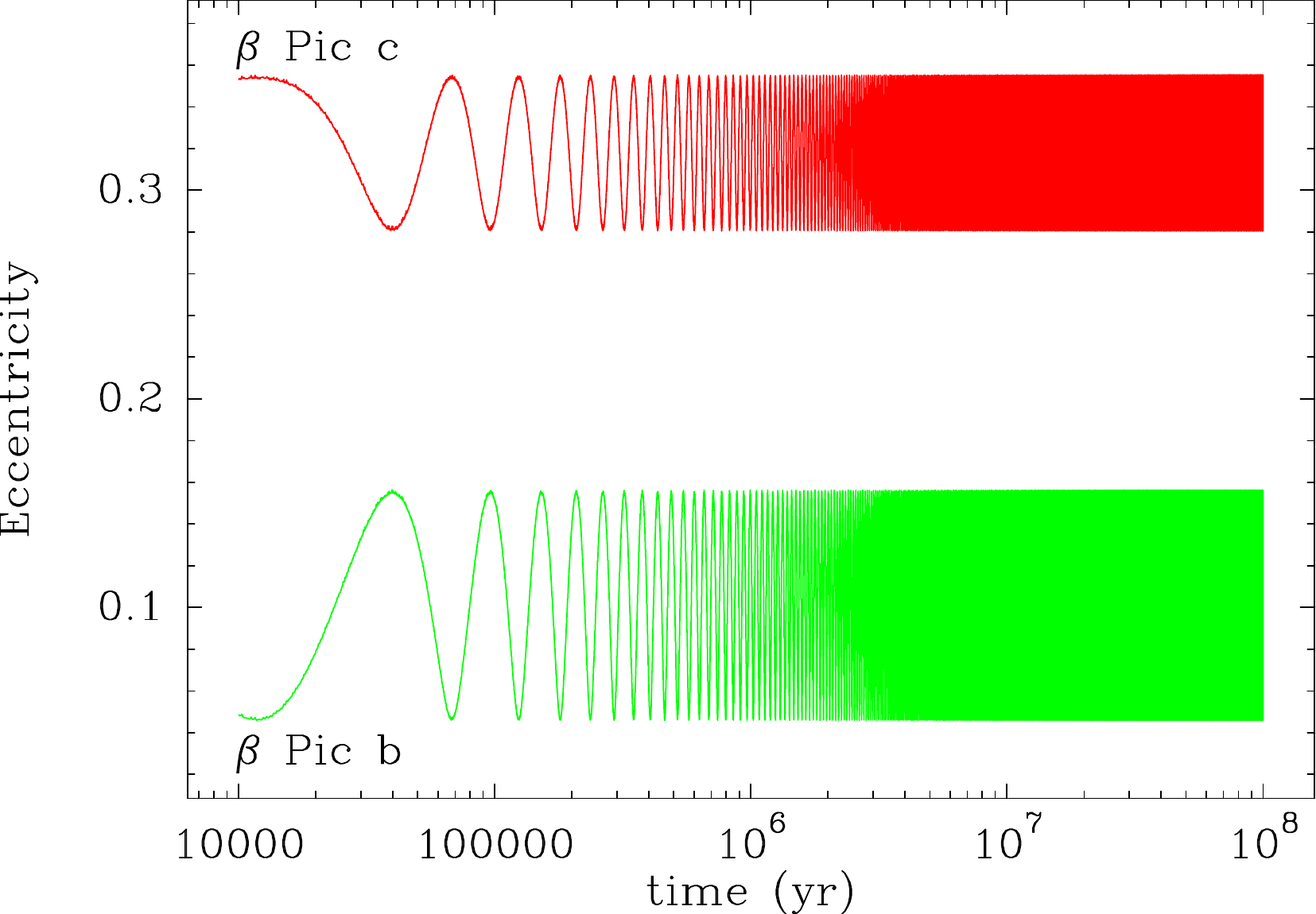} \hfil
\includegraphics[width=0.33\textwidth]{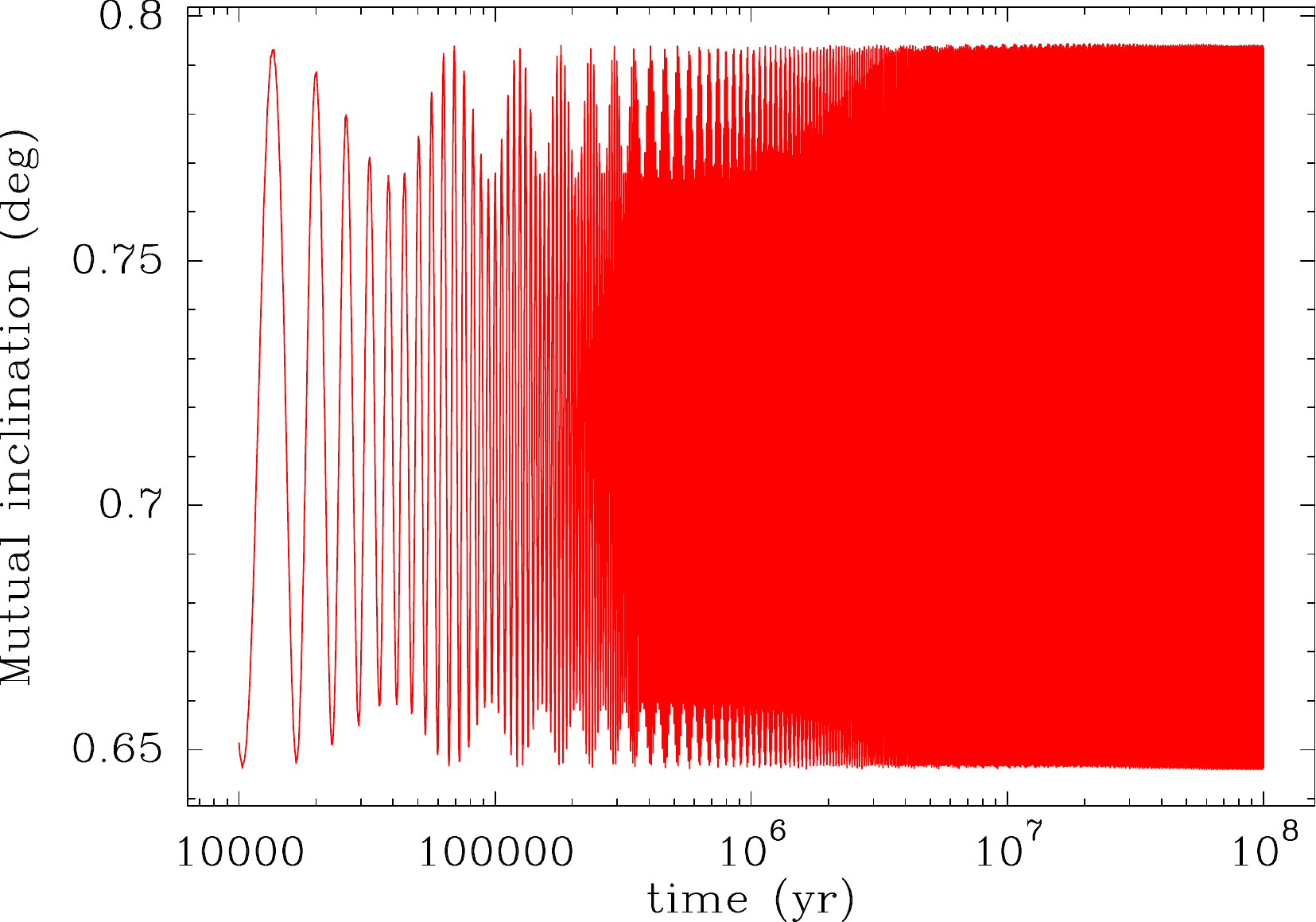}}
\caption[]{Secular evolution over $10^8\,$yr of the \bp~b+c two-planet system corresponding to solution \#2 from Table~\ref{orbits}. From left to right: Semi-major axes, eccentricities, and mutual inclination. The logarithmic time axes highlight the short-term as well as the long-term behaviour. Red always stands for \bpc\ and green for \bpb.}
\label{sec_sol2}
\end{figure*}
\begin{figure*}
\makebox[\textwidth]{
\includegraphics[width=0.33\textwidth]{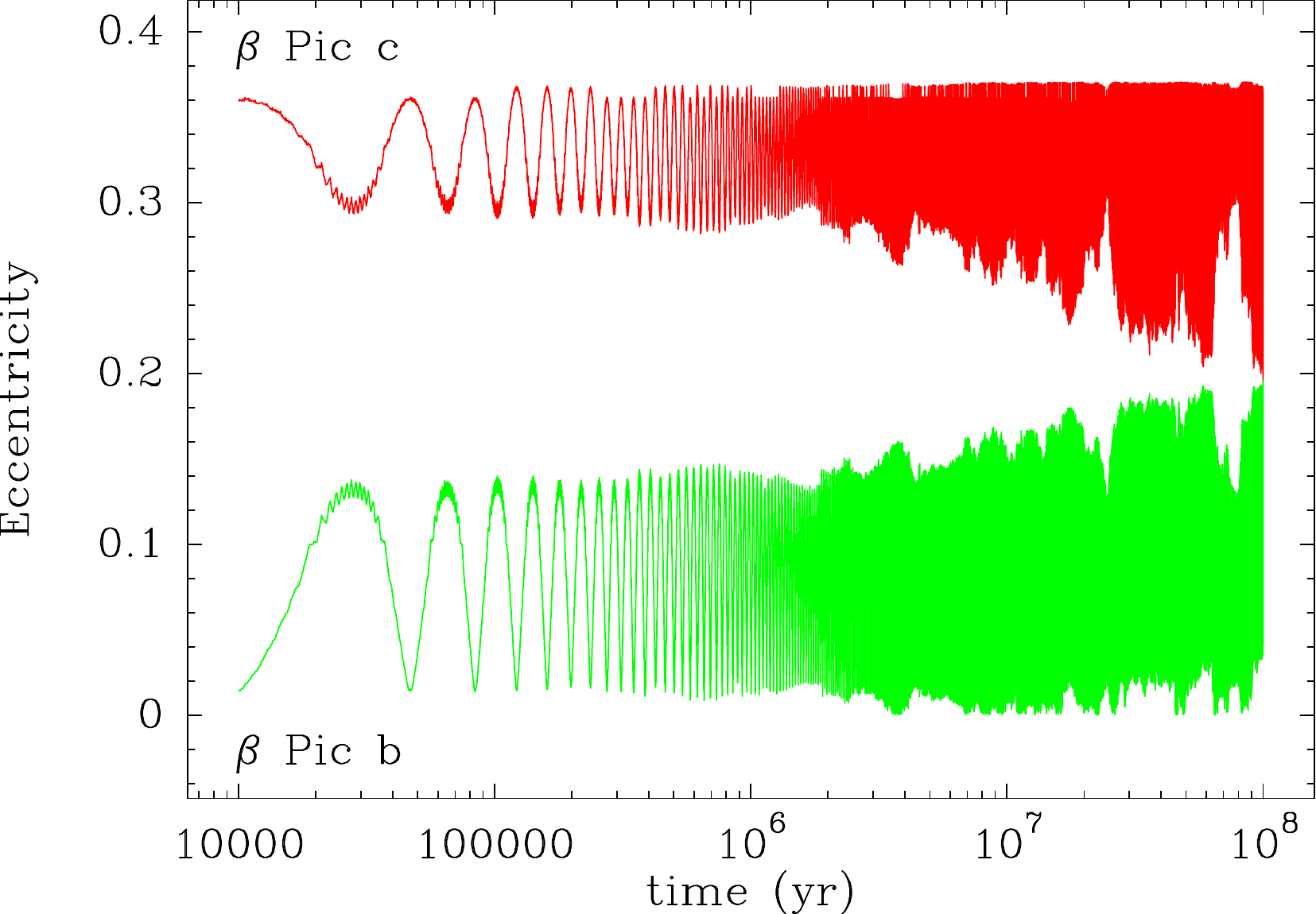} \hfil
\includegraphics[width=0.33\textwidth]{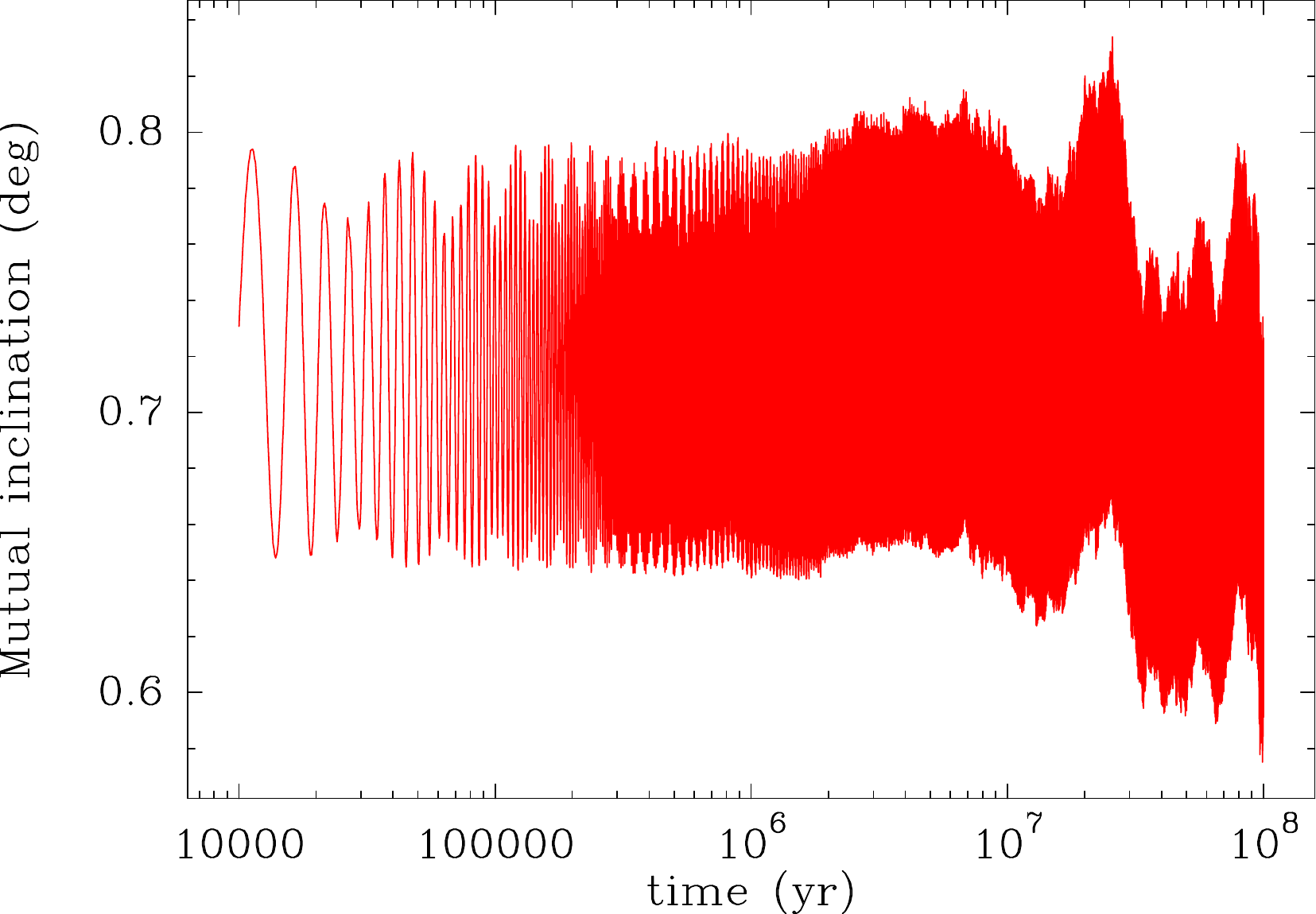} \hfil
\includegraphics[width=0.33\textwidth]{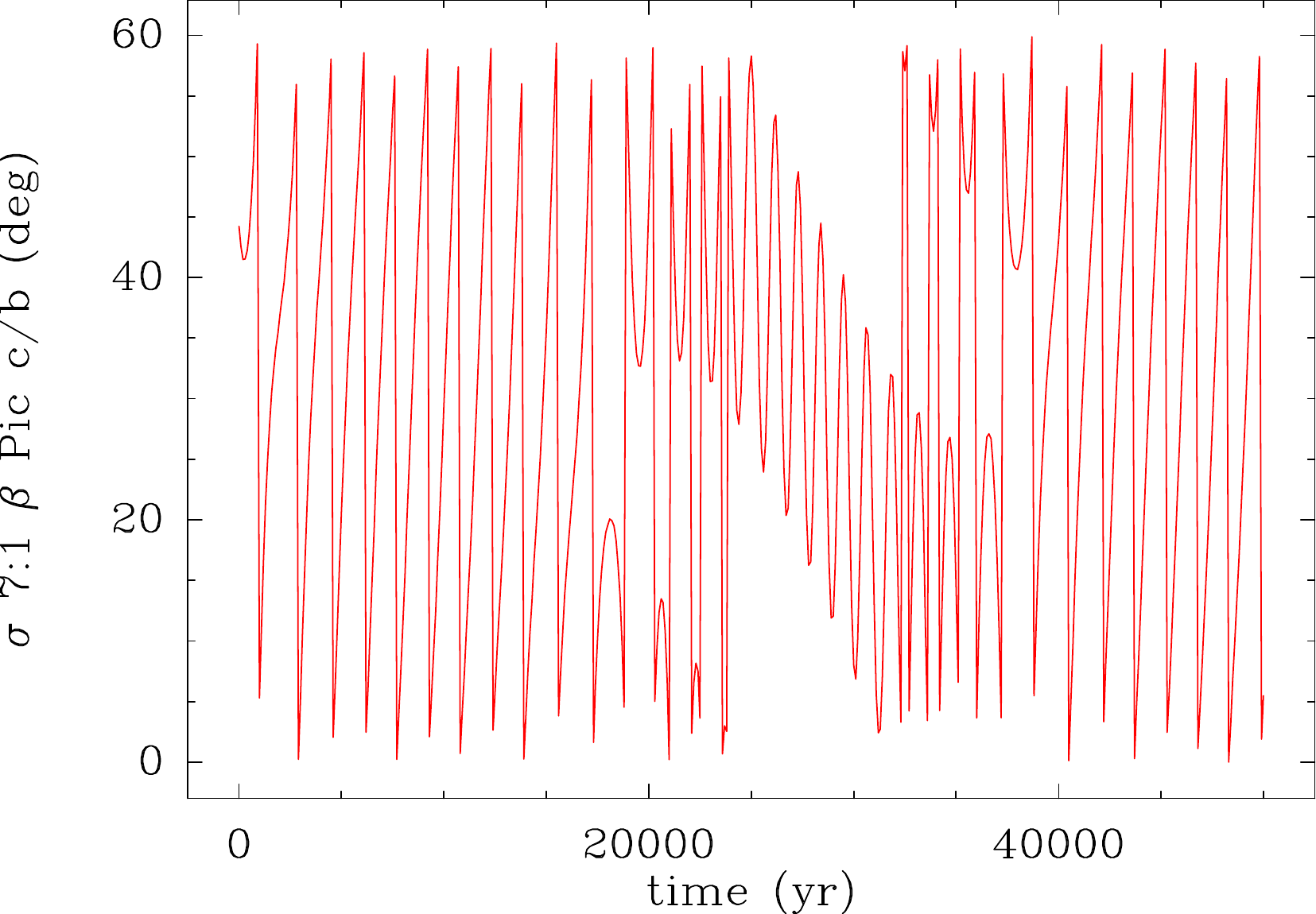}}
\vspace*{-\jot}\\
\makebox[\textwidth]{
\includegraphics[width=0.33\textwidth]{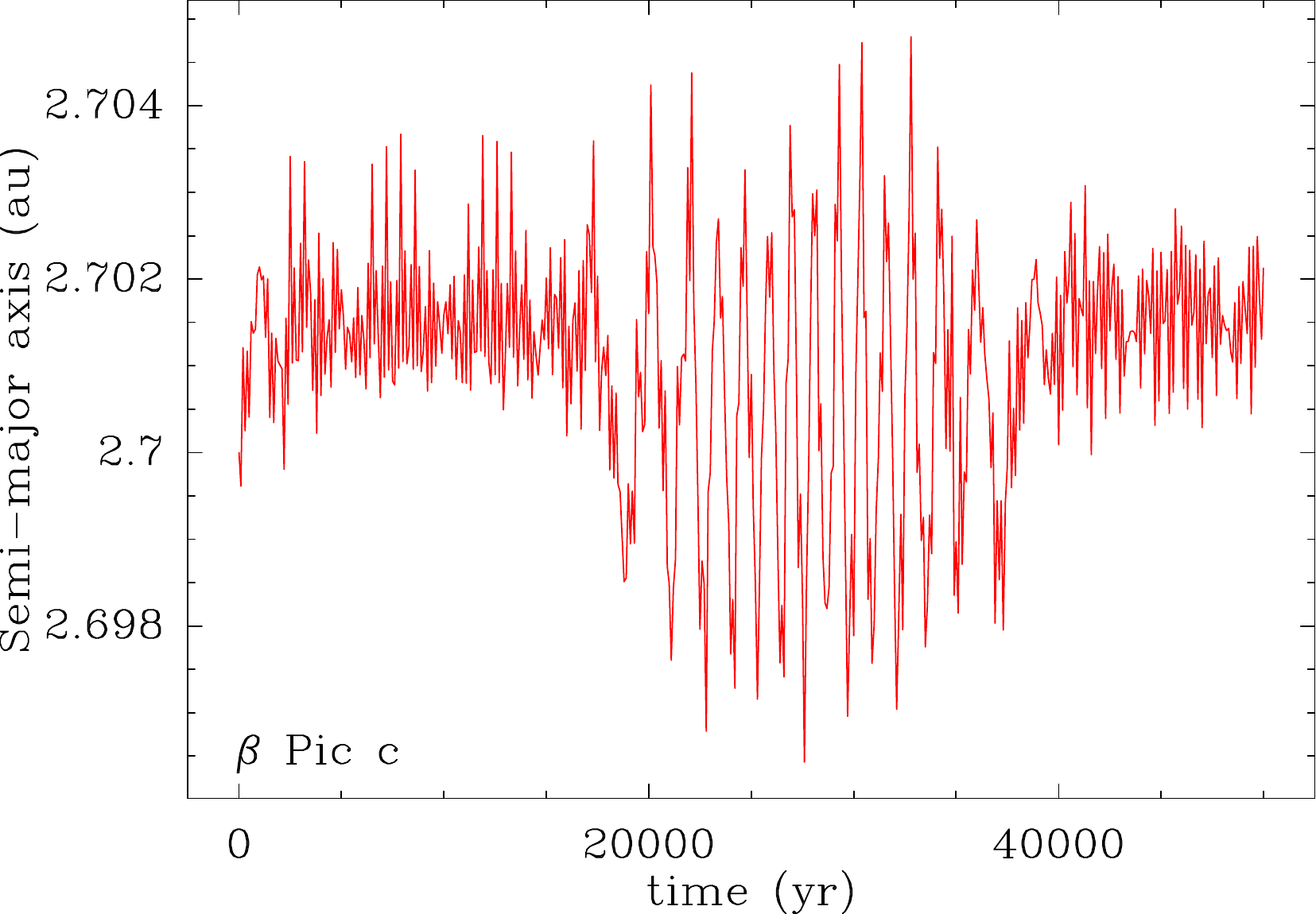} \hfil
\includegraphics[width=0.33\textwidth]{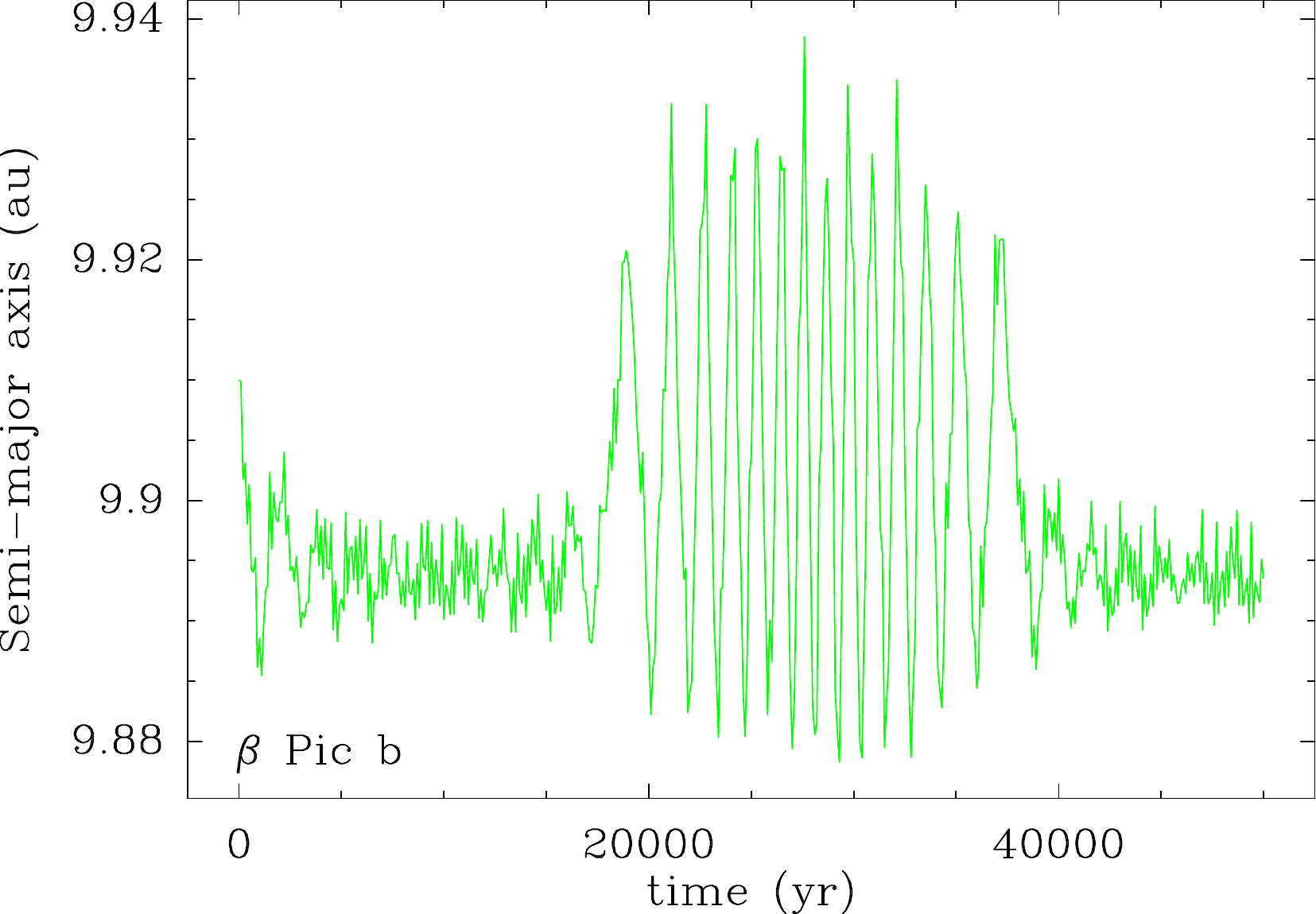} \hfil
\parbox[b]{0.30\textwidth}{\caption[]{Secular evolution over $10^8\,$yr of the \bp~b+c two-planet system corresponding to solution \#1 from Table~\ref{orbits}. Upper row, from left to right~: Eccentricities, mutual inclination, and (over the first 50,000~yr) the characteristic resonant argument $\sigma_{7:1}$ (see text). Our definition of $\sigma_{7:1}$ implies a $60\degr$ degeneracy. In our calculation, we always keep the values between 0 and $60\degr$. Hence the variations from 0 to $60\degr$ appearing here actually mean circulation of $\sigma_{7:1}$. Lower row: Short-term semi-major axis variations of the two planets over the first 50,000 yr.}\label{sec_sol1}}}
\end{figure*}
The two-planet system orbiting \bp\ is assumed to be dynamically stable. Its secular evolution is briefly mentioned in \citet{2021A&A...654L...2L}, revealing stability despite non-negligible eccentricity oscillations. Interestingly, \citet{2021A&A...654L...2L} points towards a possible 7:1 MMR between the two planets. To investigate this issue, we   integrated the system starting from the orbital solutions given by \citet{2021A&A...654L...2L}. More specifically, we integrated two configurations, listed in Table~\ref{orbits}. Solution \#1 is the best one from Table~3 of \citet{2021A&A...654L...2L} computed using all available data; solution \#2 is the one labelled `truth' from Table~B.1 of \citet{2021A&A...654L...2L}, which is basically an average of all the possible values. Many other configurations could be tested, but these can be considered representative. The two solutions are obviously close to each other, but the main difference between them appears if we compute the ratio of the two orbital periods $P_b/P_c$ in each case. We find $P_b/Pc=7.0097$ for solution \#1 and $P_b/P_c=7.466$ for solution \#2. Hence, solution \#1 is presumably locked in 7:1 MMR, whilst solution \#2 is not.

As the solutions of \citet{2021A&A...654L...2L} are given in Jacobi coordinates, the integrations were carried out over $10^8\,$yr (i.e. far above the current age of the star) using the symplectic N-body code \textsc{Swift\_hjs} \citep{2003A&A...400.1129B}, which naturally works in that coordinate system. We note that prior to all computations, all angular orbital elements have been converted relative to the invariable plane of the three-body system as a more relevant reference system. Figure~\ref{sec_sol2} shows the evolutions of the semi-major axes, the eccentricities of the two planets, and their mutual inclination for solution \#2 from Table~\ref{orbits}. We note a remarkable stability of the semi-major axes, indicating a stable system. Both eccentricities and mutual inclination exhibit the regular oscillations characteristic of a non-resonant system with regular dynamics. The longitudes of ascending nodes and arguments of periastra (not shown here) undergo very regular precession motions.

Figure~\ref{sec_sol1} shows the same evolution, but starting with solution \#1 from Table~\ref{orbits}. We do not show the general evolution of the semi-major axes as this appears identical to Fig.~\ref{sec_sol2}. We note however a more irregular evolution of the eccentricities and of the mutual inclination. This is a consequence of the 7:1 MMR. To check this point, we also show the evolution of the critical argument $\sigma_{7:1}$ of that resonance, which is defined as 
\begin{equation}
\sigma_{7:1}=\frac{7}{6}\lambda_b-\frac{1}{6}\lambda_c-\varpi_c\qquad,
\end{equation}
where $\lambda_c=$ is \bpc's mean longitudes (and similarly for \bpb) and $\varpi_c=\Omega_c+\omega_c$ is its longitude of periastron \citep[see e.g.][for a more general definition]{2016A&A...590L...2B}. The quantity $\sigma_{7:1}$ is presumably a slowly variable quantity thanks to the MMR. Resonant configurations are expected to be characterized by a libration of $\sigma_{7:1}$, whilst non-resonant ones should exhibit circulation of $\sigma_{7:1}$. 

The evolution of $\sigma_{7:1}$ over the first 50,000 years of the simulation shows a hybrid configuration. We first note a circulation up to $\sim18,000$\,yr, and then a libration motion that lasts approximately 20,000 years before the circulation starts again. What we see here is a temporary capture in 7:1 MMR. This situation appears to repeat itself all along the simulation, with a 20,000\,yr resonant capture every $\sim$40\,000\,yr. On average, the two-planet system appears to spend nearly half of its time within the resonance and half out of it. This hybrid behaviour could still be artificial and due to some numerical aliasing in Fig.~\ref{sec_sol1} as a consequence of an insufficient time resolution of the plot. This is actually not the case. Orbital data were saved every 100\,years in the simulations and plotted with the same resolution (i.e. much higher than the frequency of the $\sigma_{7:1}$ oscillations).

Moreover, the hybrid behaviour is not specific to the configuration tested. We checked many other solutions close to solution \#1 from Table~\ref{orbits}, in particular changing \bpc's semi-major axis to match the exact 7:1 MMR value. We found no configuration where the system permanently remains locked in the MMR. In some cases $\sigma_{7:1}$ always circulates, which denotes a non-resonant configuration, and thus similar to solution \#2. In all other cases we have temporary resonant captures as in Fig.~\ref{sec_sol1}, so that the two secular evolutions we detail here are representative for all possible situations.

In Fig.~\ref{sec_sol1} we also show the detailed evolution of the semi-major axes of the two planets over the first 50,000 years. We note rapid very small amplitude changes when the system is out of the 7:1 MMR, and more pronounced oscillations in the resonant phase. This behaviour is characteristic of a temporary resonant capture. Out of any MMR, the semi-major axes undergo no secular variations apart from minor short term changes. These rapid changes would also show up in Fig.~\ref{sec_sol2} if we focused on the detailed evolution of each semi-major axis. Conversely, resonant configurations are known to generate secular oscillations of semi-major axes coupled with eccentricity and $\sigma$ oscillations \citep{2016A&A...590L...2B}. This is exactly what we observe in Fig.~\ref{sec_sol2}.

Hence, we confirm that solution \#1 corresponds to a mode of regular temporary captures of the two-planet system in 7:1 MMR. Given the high masses of the two planets, this resonance (a sixth-order MMR) is actually too weak to sustain itself permanently.
\section{Simulations with a disk of planetesimals}
\label{simdisk}
\subsection{Global simulation}
\begin{figure*}[tbp]
\makebox[\textwidth]{
\includegraphics[width=0.49\textwidth]{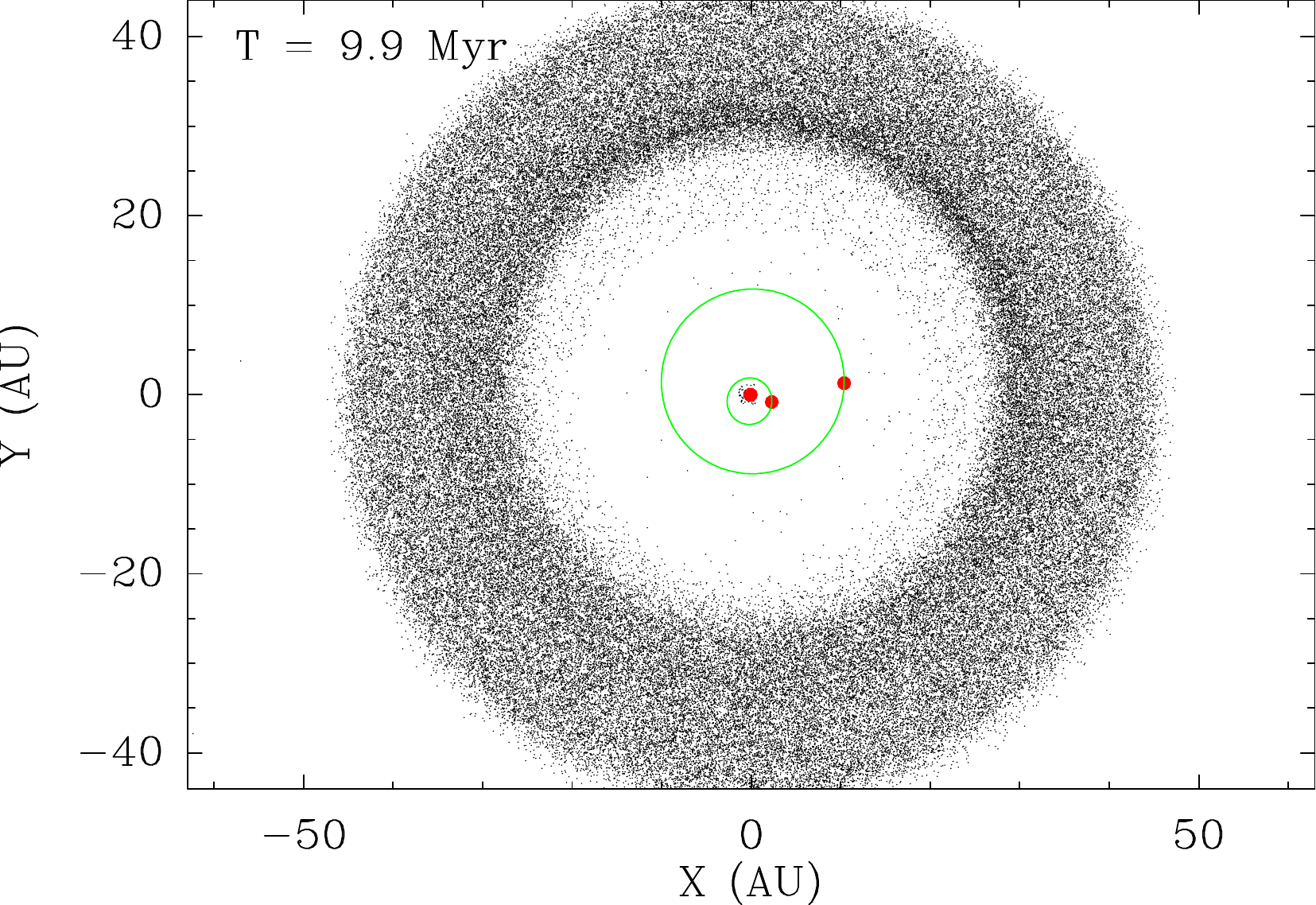} \hfil
\includegraphics[width=0.49\textwidth]{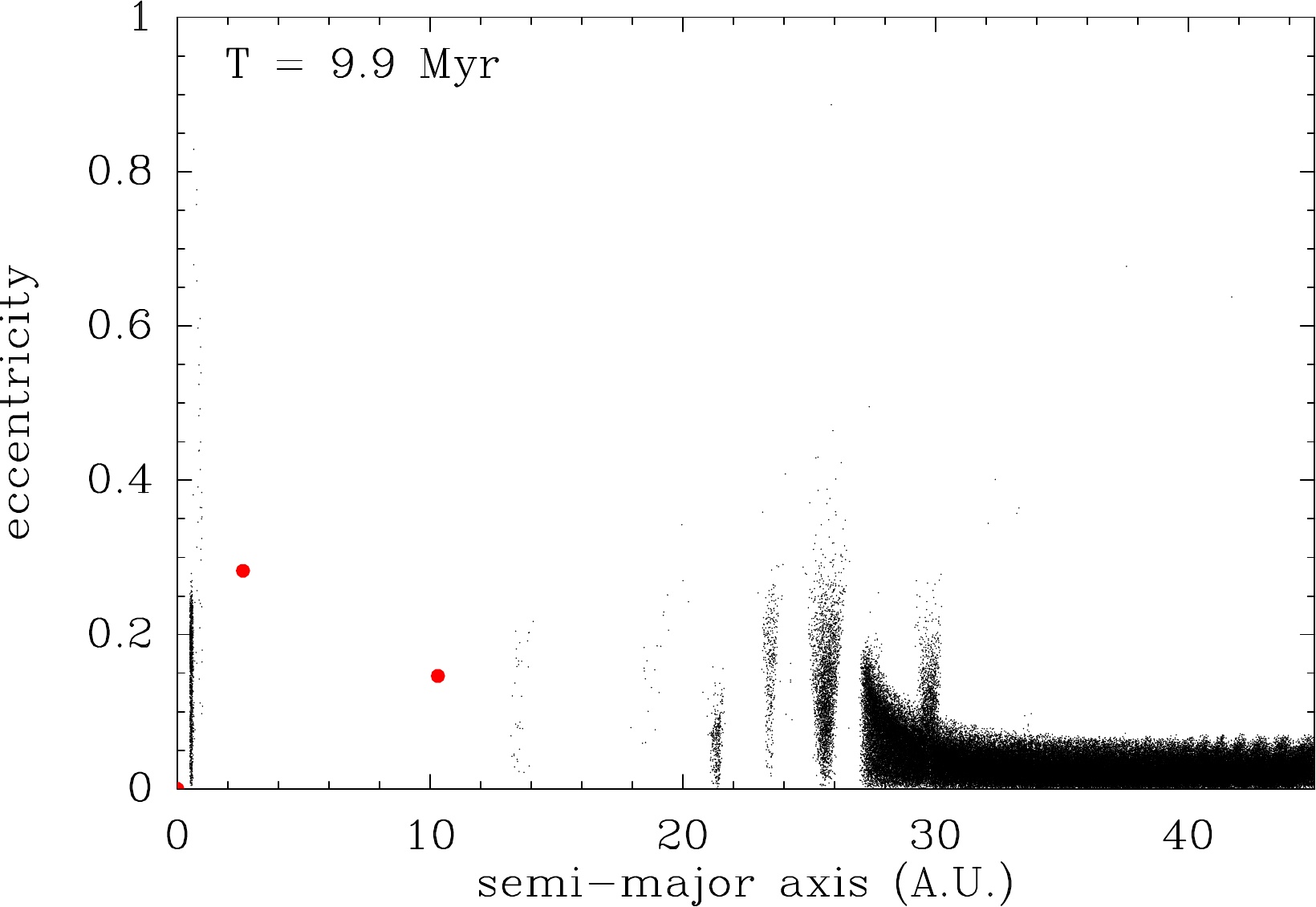}}
\caption[]{Outcome after a 10\,Myr simulation of a disk of planetesimals initially between 0.5\,au and 45\,au in the environment of \bpb\ and \bpc\ taken from solution \#2 from Table~\ref{orbits}, viewed from above (left), and displayed in $(a,e)$ space (right). The red bullets show the location of the planets and the star, and the black dots are the remaining planetesimals.} 
\label{lac_truth_10myr}
\end{figure*}
\begin{figure*}[tbp]
\makebox[\textwidth]{
\includegraphics[width=0.49\textwidth]{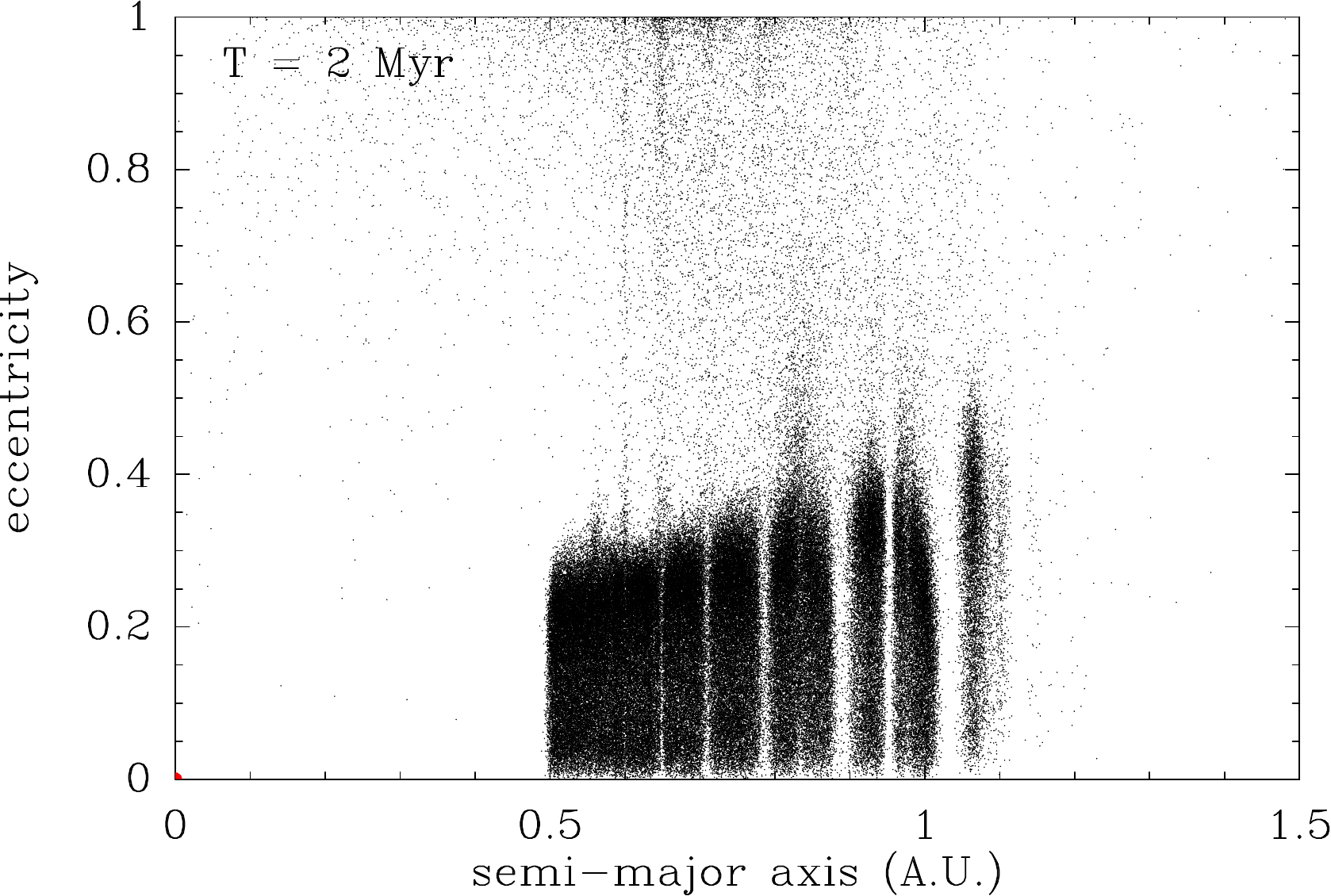} \hfil
\includegraphics[width=0.49\textwidth]{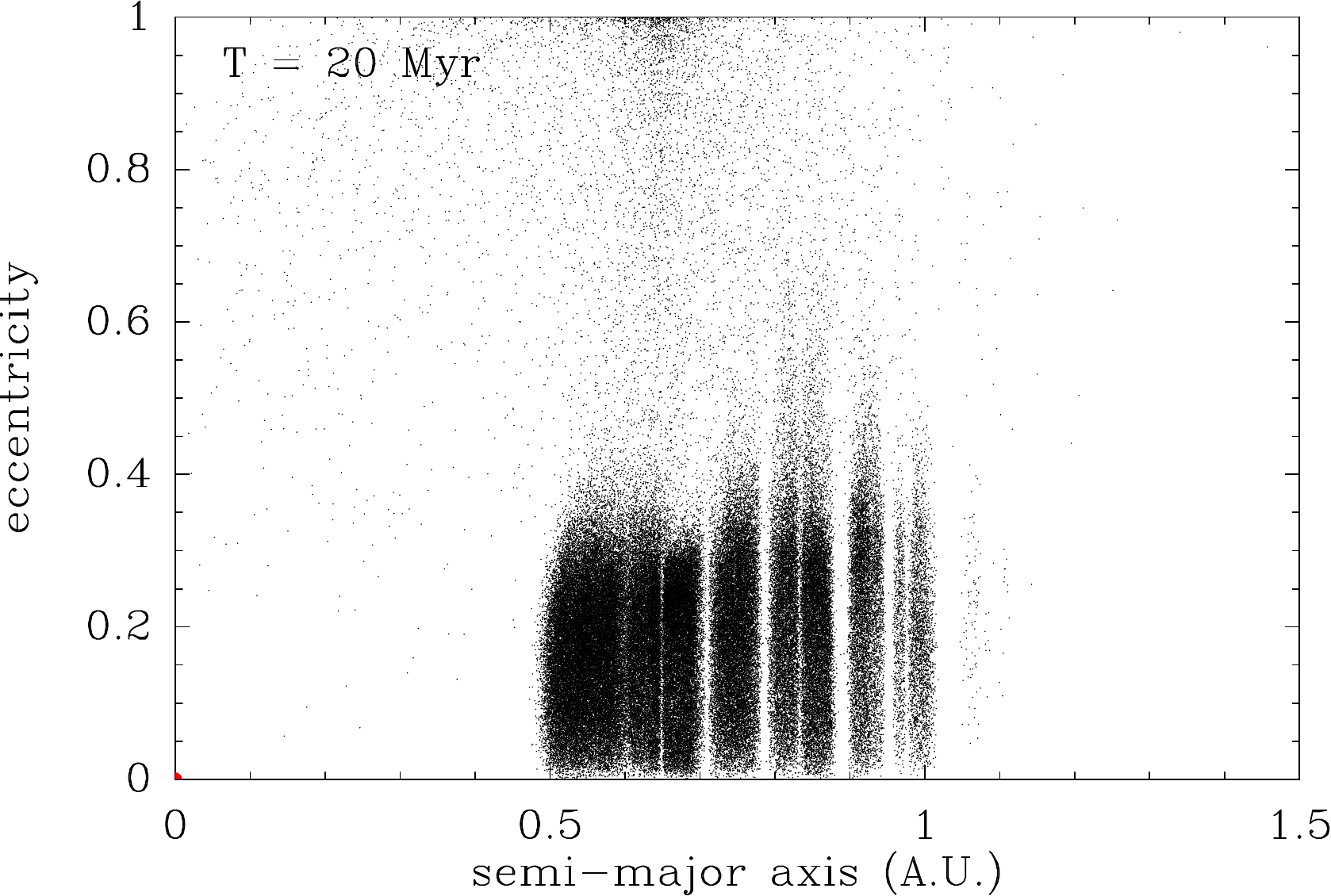}}
\caption[]{Outcome in $(a,e)$ space of a disk of planetesimals initially between 0.5\,au and 1.5\,au, under the same conditions as in Fig.~\ref{lac_truth_10myr}, after 2\,Myr (left) and 20\,Myr (right). Both planets are outside the panel.} 
\label{lac_truthc}
\end{figure*}
\begin{figure*}[tbp]
\makebox[\textwidth]{
\includegraphics[width=0.49\textwidth]{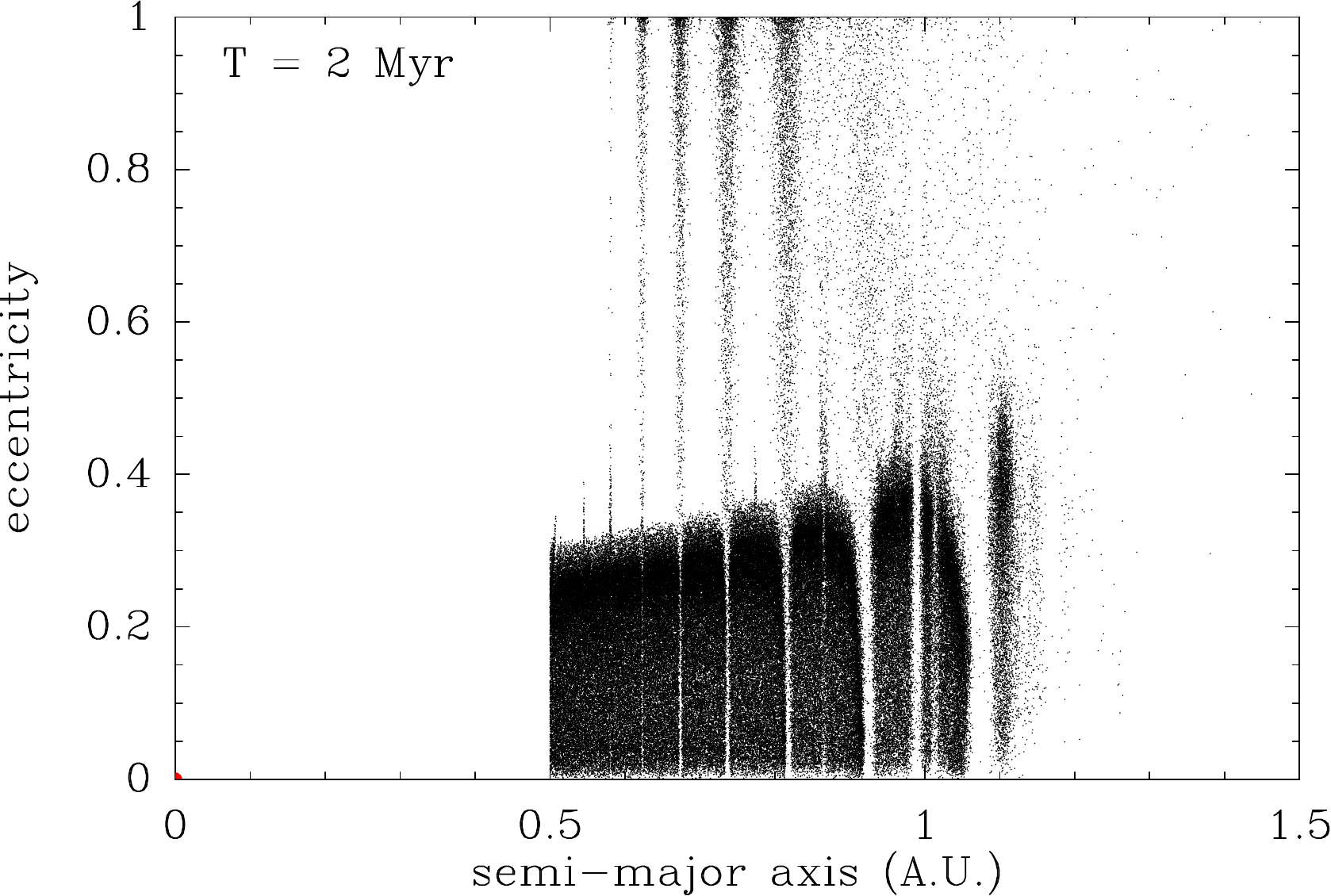} \hfil
\includegraphics[width=0.49\textwidth]{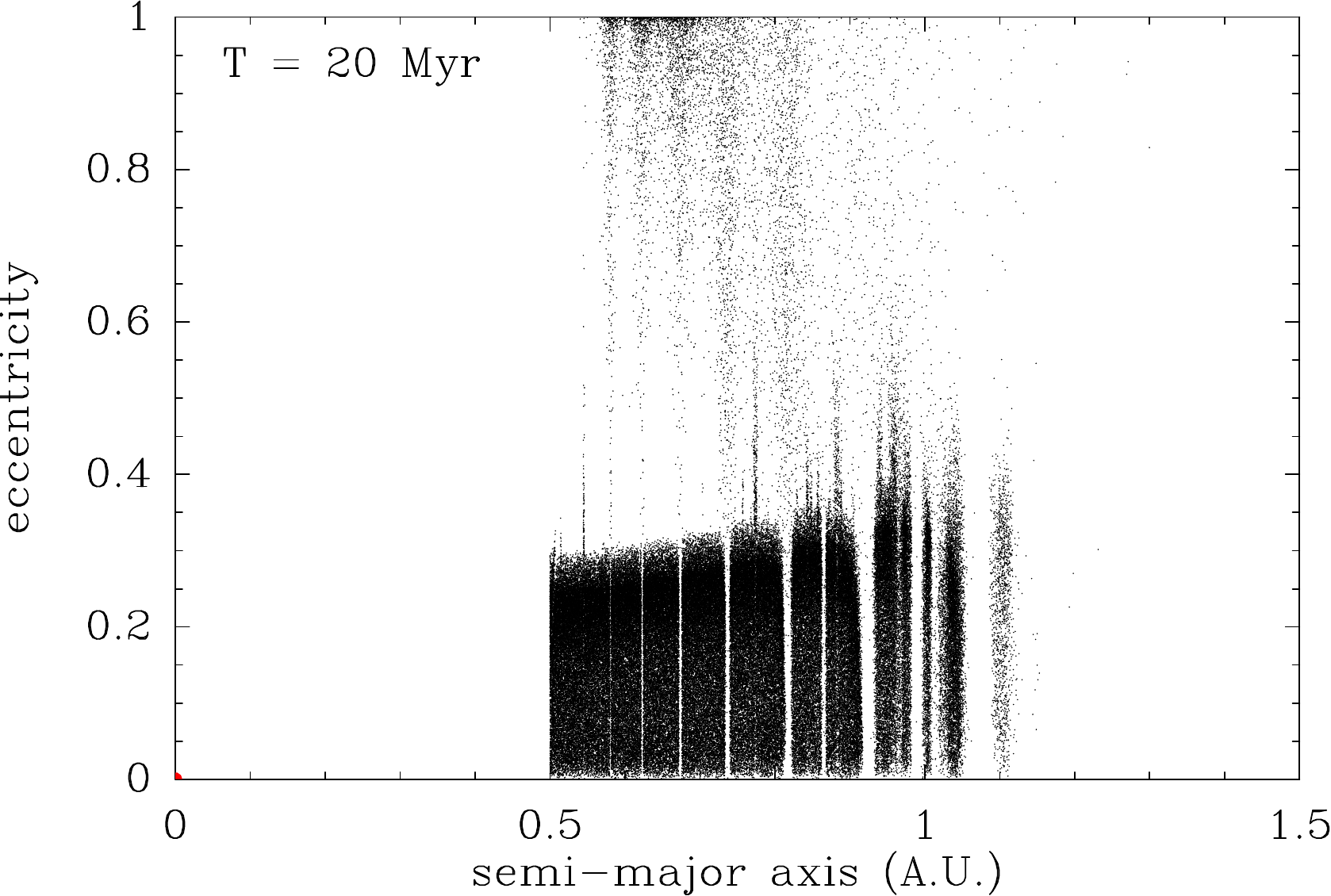}}
\caption[]{Same as Fig.~\ref{lac_truthc}, with the same initial disk of particles between 0.5\,au and 1.5\,au, but starting from solution \#1 from Table~\ref{orbits} for \bpb\ and \bpc.}
\label{lac_rvc}
\end{figure*}
In this section we deal with presenting simulations involving the two planets \bpb\ and \bpc\ and a disk of planetesimals. The main goal is to investigate where planetesimals can dynamically survive in this environment, and whether the suspected FEB-generating mechanism is still active. The subsequent simulations assume a starting disk of 400,000 planetesimals treated as massless particles with initial semi-major axes randomly drawn between given bounds that depend on the simulation. Eccentricities are drawn between 0 and 0.05, and inclinations between 0 and $2\degr$ with respect to the invariable plane of the two-planet planetary system. Other orbital elements such as longitudes of ascending nodes, longitudes of periastra,  and initial mean longitudes are randomly drawn between 0 and $360\degr$. Computations are carried out using the \textsc{Rmvs3} symplectic integrator \citep{1994Icar..108...18L}, a version of the initial mixed-variable symplectic integrator \citep{1991AJ....102.1528W} that better treats close encounters. We note that from a technical point of view, \textsc{Rmvs3} works in heliocentric coordinates, whilst the orbits listed in Table~\ref{orbits} are given in Jacobi coordinates. Hence, prior to launching any simulation, we first convert the planetary orbital elements of Table~\ref{orbits} to heliocentric. This actually only affects \bpb, as for \bpc\ (the innermost planet) the two sets of elements coincide; however,  even for \bpb\ the difference is small as the stellar mass dominates.

Figure~\ref{lac_truth_10myr} shows the result of such a simulation after 10\,Myr assuming orbital  solution \#2 (non-resonant) from Table~\ref{orbits} for the planets, and a disk of planetesimals ranging initially between 0.5\,au and 45\,au. The remaining disk is shown as a planar upper view and in (semi-major axis $a$, eccentricity $e$; herefter $a,e$) space. The planetesimals appear as black dots. The disk appears strongly depleted interior to $\sim 25\,$au. Most of the  planetesimals initially located closer in have been removed by perturbations and close encounters due to the planets. The inner edge of the disk appears strongly carved by various MMRs with \bpb as all subsisting rings inside $\sim 27\,$au actually correspond to these MMRs: the 2:3 and 2:5 at $a=13.03\,$au and $a=18.33\,$au (marginally); the 1:3 at $a=20.7\,$au, the 2:7 at $a=22.94\,$au, and the 1:4 at $a=25.07\,$au. Particles trapped in these resonances subsist here thanks to a phase-protection mechanism that prevents close encounters with \bpb. The resonant dynamics excites the eccentricities of these particles, so that their mean eccentricity is significantly larger than those from the main surviving non-resonant disk beyond 25\,au. This particularly shows up in the $(a,e)$ map of Fig.~\ref{lac_truth_10myr}. A ring of more eccentric particles, actually corresponding to the 1:5 MMR with \bpb\ at $a=29.09\,$au, even shows up in the middle of the unperturbed disk.

It is important to note that no particle was able to survive between the two planets. In the context of the FEB-generating mechanism by MMRs with \bpb, this is problematic as the model described by \citet{1996Icar..120..358B,2000Icar..143..170B} and \citet{2001A&A...376..621T} pointed out the major inner (4:1, 3:1) MMRs with that planet as their probable source. These MMRs all fall in the region between \bpb\ and \bpc. For instance, the 4:1 MMR, which appeared to be the most powerful source of FEBs, is located close to 4\,au where no planetesimal appears dynamically stable over 10\,Myr (and actually much less). It thus turns out that even if \bpb\ appears to almost perfectly match the giant planet suspected more than 20 years ago to be responsible for the FEB mechanism, the presence of \bpc\ orbiting inside invalidates this picture.

The outer MMRs with \bpb\ quoted above that carve the outer planetesimal disk could not also constitute valuable sources of FEBs. It actually shows up in Fig.~\ref{lac_truth_10myr} that even if the eccentricity of the resonant particles is enhanced by the resonant mechanism, none of them is able to reach a high enough eccentricity to become a FEB ($e\ga 0.96$ would be required here). This may also be due to the presence of \bpc\ orbiting inside that acts as a barrier.
%
\subsection{Focus on the inner ring}
\label{inner}
\subsubsection{Technical issue}
In Fig.~\ref{lac_truth_10myr}, however, another ring of surviving particles is present inner to \bpc's orbit. We focus here on this inner ring. To achieve a better spatial resolution in this area, a new simulation was initiated, now taking  400,000 particles between 0.5\,au and 1.5\,au, as the previous simulation had revealed that no particle was able to survive beyond 1.5\,au due to the presence of \bpc. The inner bound of 0.5\,au was fixed according to the evaporation limit of refractory compounds determined by \citet{1998A&A...338.1015B}. Hence no planetesimal was expected to sustainably orbit the star closer to this threshold. This is of course an approximate limit that depends actually on the real composition of the bodies under consideration. \citet{1998A&A...338.1015B} shows for instance that more carbonaceous FEBs may be able to survive closer to the star. The simulation was also extended up to 20\,Myr (i.e. the present-day estimated age of \bp).

Technically speaking, this new simulation was initially carried out with the same integrator (\textsc{Rmvs3}) as before, but although the results were satisfying at first glance, they also appeared inaccurate. Many particles around $\sim 1\,$au underwent a secular drift in semi-major axis that seemed spurious. Other attempts with reduced  time-steps revealed that this secular drift actually depends on the time-step, proving its numerical origin. Nonetheless, it was not possible to find any convenient time-step value able to eliminate this effect. After analysis, the origin of this phenomenon was identified as inherent to the way close encounters are treated in \textsc{Rmvs3}. Whenever a test particle gets sufficiently close to a massive planet, \textsc{Rmvs3} automatically reduces the time-step for that particle during the encounter to better resolve it. This temporarily breaks the symplecticity of the integration, but does not affect the global relevance of the integration, as individual particles undergo close encounters only occasionally. This statement applies to the simulation described in Fig.~\ref{lac_truth_10myr}, in particular for the outer part of the disk. Due to the high mass of \bpc\ and to its fairly high eccentricity, particles orbiting around $\sim 1\,$au regularly get sufficiently close to \bpc\ (near conjunction) to be considered as having a close encounter with it by \textsc{Rmvs3} without being ejected. However, as this occurs quite often, the integration for such particles is never symplectic. This is the origin of the semi-major axis drift observed.

To overcome these difficulties, we changed the integrator. We note that adopting other integrators that also temporarily break the symplecticity of the integration, such as \textsc{Mercury} \citep{1999MNRAS.304..793C} and \textsc{Rebound} \citep{2012A&A...537A.128R}, would not help much more. We tried \textsc{Symba} \citep{1998AJ....116.2067D}, a more sophisticated integrator that remains strictly symplectic while handling close encounters. This also turned out to be inaccurate, but for another reason. Due to the use of the democratic heliocentric method (DH), \textsc{Symba} hardly resolves orbits with very small periastron, which is exactly what we have here with FEBs. Therefore, the computation became extremely long, although the semi-major axis drifts were no longer present. Conversely, the classical second-order Wisdom--Holman (WH) mapping naturally handles such orbits, but is not able to compute close encounters. We finally moved to a higher-order symplectic integrator than WH with no change in time-step during close encounters, and modified WH accordingly. We adopted the sixth-order S6B scheme described by \citet{2000AJ....119..425C}, assuming a time-step of $1.33\times10^{-2}\,$yr (i.e. 1/20 of the smallest orbital period). This way semi-major axis drifts were no longer present, and thus we can trust our integrations.
\subsubsection{Results}
Figure~\ref{lac_truthc} presents the result of this new simulation, a view of the inner ring in $(a,e)$ space after 2\,Myr and 20\,Myr. It appears severely carved down to $\sim 1\,$au by perturbations arising mainly from \bpc\ located at 2.6\,au. Up to $\sim 1\,$au, most disk particles appear to still be present after 20\,Myr, but many of them have reached high eccentricity values, thus becoming FEB candidates. We note that for a planetesimal starting at $\sim 0.8\,$au, reaching the FEB state with a periastron $\la 0.4\,$au only requires an eccentricity $e\ga 0.5$, which is easily realized for many particles here. As can be seen in Fig.~\ref{lac_truthc}, the eccentricity increase is preferably done along vertical lines (i.e. at specific semi-major axis values) that can be easily identified with many inner MMRs with \bpc, for example the 4:1 at $a=1.030\,$au, the 9:2 at $a=0.952\,$au, the 5:1 at $a=0.888\,$au, the 6:1 at $a=0.786\,$au, the 7:1 at $a=0.709\,$au, the 8:1 at $a=0.649\,$au, the 9:1 at $a=0.600\,$au. Figure~\ref{ae_bpicbc_evap} displays the location of all these MMRs. We note that some very high-order MMRs, such as 13:2 and 16:3, are active FEB sources. The process is highly active at $t=2\,$Myr, but it is still present at $t=20\,$Myr.

Figure~\ref{lac_rvc}, to be compared to Fig.~\ref{lac_truthc}, presents the result of another simulation, but where solution \#1 from Table~\ref{orbits} (partially resonant) for the initial orbital configurations of \bpb\ and \bpc\ is assumed instead of solution \#2 in Fig.~\ref{lac_truthc}. There are actually minor differences between the two simulations, but in both cases the disk appears carved with very similar shapes down to 1--1.1\,au. The number of particles reaching high eccentricities at 20\,Myr is comparable, with slight differences in the distribution between different MMRs. The disk appears slightly more carved close to 1\,au in the non-resonant case. This is due to the semi-major axis difference for \bpc\ between the two solutions (Table~\ref{orbits}), which causes a more important carving in the non-resonant configuration. In the resonant case as well the resonances appear somewhat more distinctly. Both simulations nevertheless show that MMRs with \bpc\ constitute potential sources of FEBs.

This study drives us towards a new picture of the FEB generation mechanism in the \bp\ disk. MMRs are still the preferred route as, like Kozai resonance, they can constitute a long-term source of stargrazers and, contrary to Kozai resonance, they trigger an asymmetric infall \citep{1996Icar..120..358B,2000Icar..143..170B,2001A&A...376..621T} that better matches the observational statistics \citep{1996A&A...310..181B}. We were previously considering inner MMRs with a single giant planet similar to \bpb, with FEB progenitors starting around 4--5\,au. The presence of \bpc\ invalidates this picture. We are now considering planetesimals trapped in MMRs with \bpc, thus starting from a region much further in, between $\sim 0.5\,$au and $\sim 1\,$au. Moreover, the MMRs involved in the previous model were mainly 4:1, 3:1, and 5:2. We consider  here much higher-order MMRs with \bpc, for example 7:1 and 8:1. This is due to \bpc's high mass and eccentricity. Under such conditions, lower-order MMRs located further out, and thus closer to \bpc\ fall in the chaotic unstable region. The 4:1 MMR, located slightly outside 1\,au at the outer edge of the surviving disk of planetesimals, is marginally active at 2\,Myr, but no longer at 20\,Myr. Conversely, higher-order MMRs located closer to the star appear now as sources of FEBs thanks to \bpc's high eccentricity. All our previous studies \citep{1996Icar..120..358B,2000Icar..143..170B,2001A&A...376..621T,2007A&A...466..201B} considered a low-eccentricity perturber with $e\la 0.1$. \citet{2017A&A...605A..23P} showed that the mechanism remains active in higher-eccentricity regimes. Actually, the higher the planet's eccentricity, the more efficient the mechanism. The high-order MMRs considered here do not constitute efficient sources of FEBs for $e\la 0.1$, but do with $e=0.33$, as for \bpc\ here.
\subsection{Taking evaporation into account}
\label{evap}
\begin{figure*}[tbp]
\makebox[\textwidth]{
\includegraphics[width=0.49\textwidth]{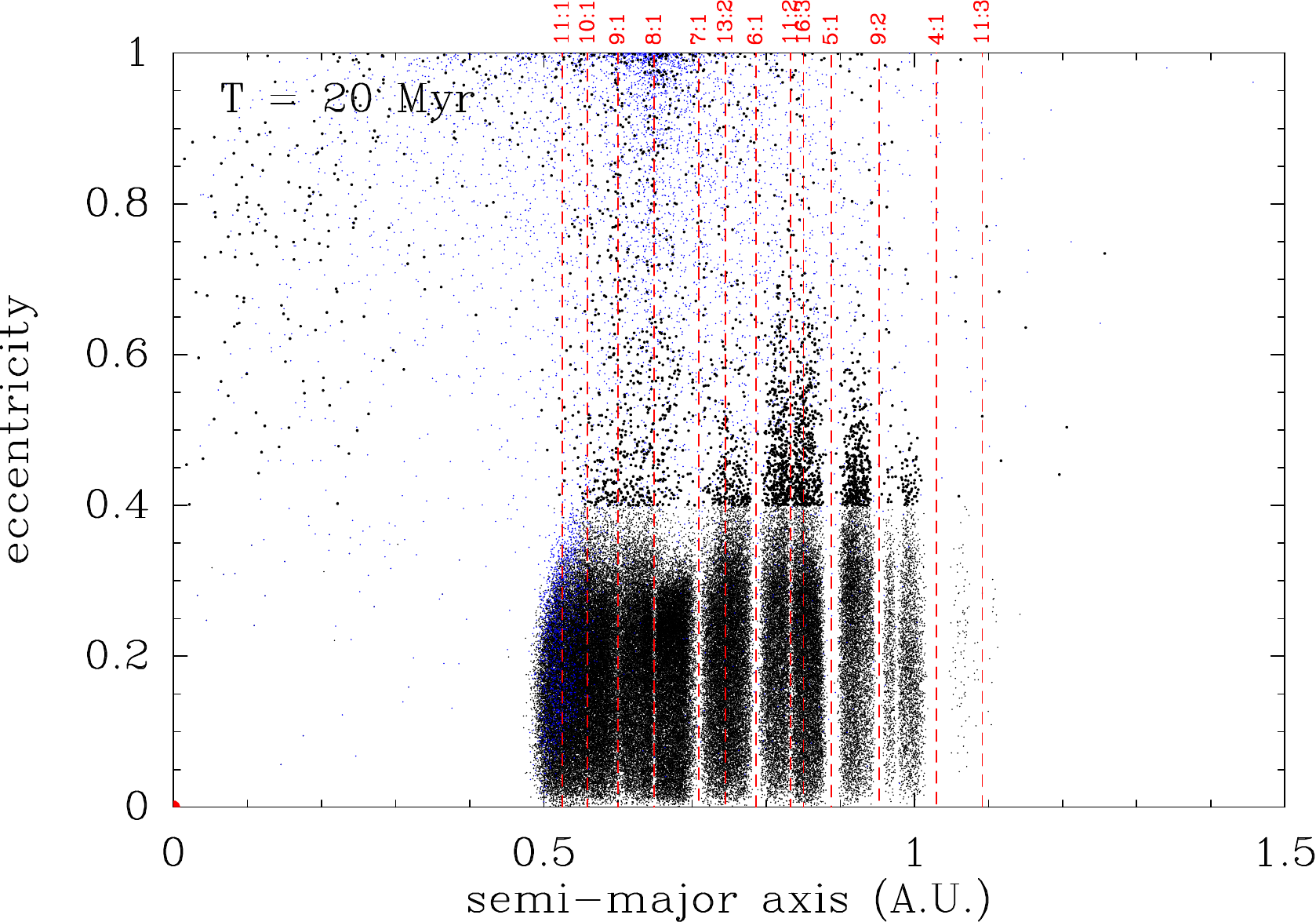} \hfil
\includegraphics[width=0.49\textwidth]{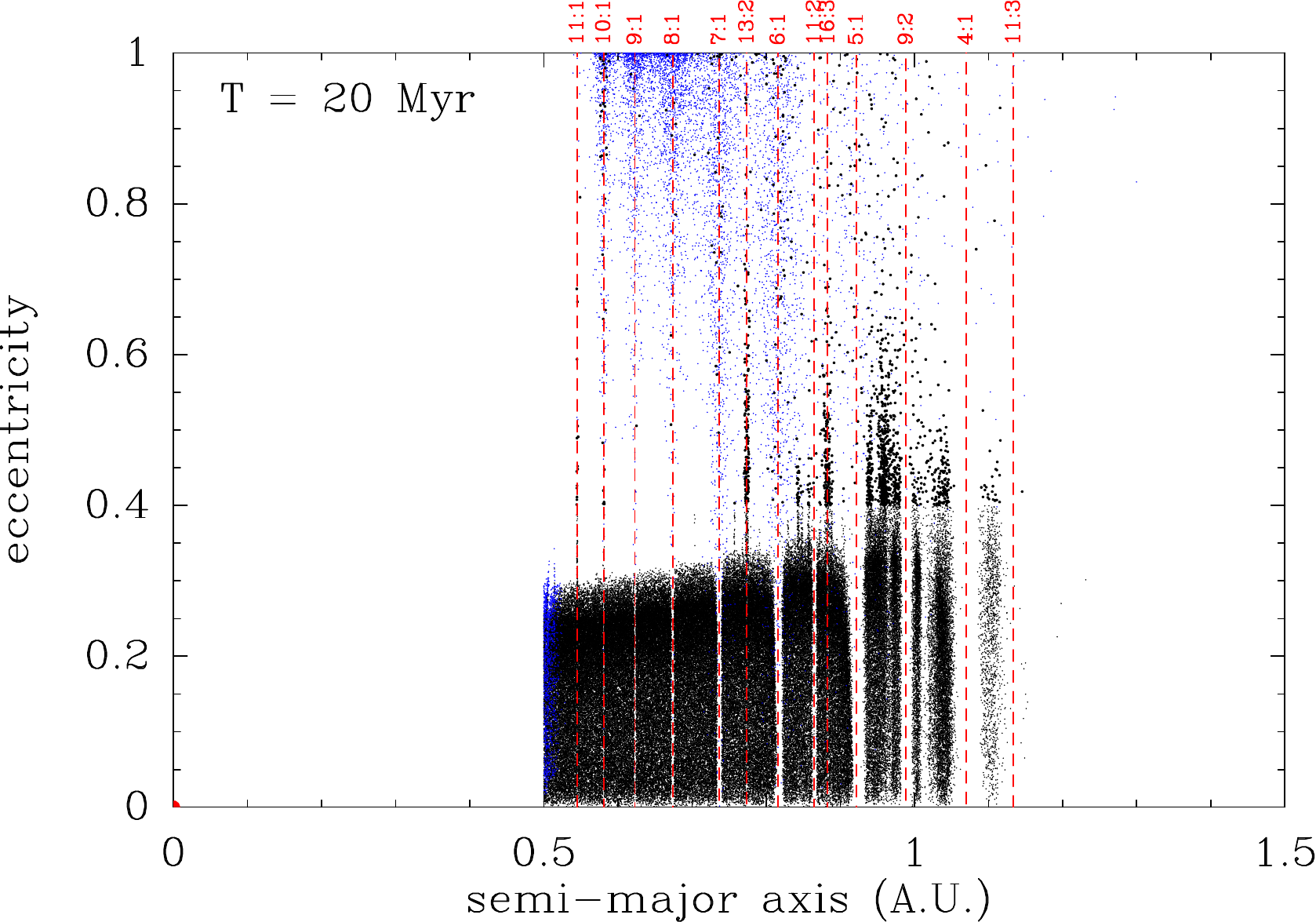}}
\caption[]{Same as Figs.~\ref{lac_truthc} and \ref{lac_rvc} at $t=20\,$Myr, but taking into account gradual evaporation of FEBs, with solution \#2 (left plot) and \#1 (right plot) from Table~\ref{orbits}. All particles appearing in blue are virtually already evaporated and should not account for the FEB statistics at that time. The locations of the main MMRs with \bpc\ are indicated with red vertical bars.}
\label{ae_bpicbc_evap}
\end{figure*}
Figures~\ref{lac_truthc} and \ref{lac_rvc} could appear misleading as they actually show potential FEB candidates moving in MMRs with \bpc\ at $t=2\,$Myr and at $t=20$\,Myr, but as long as bodies get trapped in the resonant eccentricity increase process and get closer to the star, they start to evaporate and may be quickly fully destroyed. Evaporation was not taken into account in our dynamical simulation as this would have resulted in an excessively long computation time and, in first approximation, the evaporation process does not affect the dynamics. It would nevertheless be relevant to determine whether particles moving in MMRs at 20\,Myr in the preceding simulations are virtually already evaporated or not as they may account for FEB statistics in the latter case only. 

Computing gradual evaporation of FEBs along successive periastron passages is not straightforward as this may depend highly on its chemical composition, porosity, and other parameters. To date, the most complete study of evaporation process of exocomets in the vicinity of \bp\ has been made by \citet{2003A&A...409..347K}. It shows that, on average, most of the sublimation of refractories occurs closer to $\sim0.2\,$au. The process actually starts below $\sim0.4\,$au \citep{1998A&A...338.1015B}. To simplify the treatment, we assume a surface evaporation rate $\propto 1/r^{16}$ for $r<0.4\,$au. This actually represents the best power-law fit of the production rate curves of \citet{2003A&A...409..347K} as long as the simulated FEBs are not too   destroyed (i.e. for $r\ga 0.2\,$au). More specifically, we consider a reference mass evaporation rate $z_0$ for a reference body of radius $s_0=10\,$km at reference distance $r_0=0.2\,$au,  defined as
  \begin{equation}
    \alpha=\frac{z_0r_0^{16}}{4\pi s_0^2}\qquad.
  \end{equation}
  Then, the mass loss rate of any body of radius $s$ at distance $r$ reads
  \begin{equation}
    \frac{\rd m}{\rd t}=\frac{4\pi\alpha s^2}{r^{16}}\qquad,
  \end{equation}
  and the radius decrease of the body is
  \begin{equation}
    \frac{\rd s}{\rd t}=\frac{1}{4\pi s^2\rho}\,\frac{\rd m}{\rd t}
    =\frac{\alpha}{\rho r^{16}}\qquad,
  \end{equation}
  where $\rho$ is the bulk density of the material. Integrating this expression over one periastron passage, we get
  \begin{equation}
    \Delta s=2\int_0^{\nu_\mathrm{max}}\frac{\rd s}{\rd t}\times\frac{\rd\nu}{\rd\nu/\rd t}\qquad,
  \end{equation}
  where $\Delta s$ is the full radius loss over one periastron passage, $\nu$ is the true anomaly along the orbital path, and $\nu_\mathrm{max}$ is the value of $\nu$ corresponding to $r=r_\mathrm{max}=0.4\,$au, the maximum evaporation distance of refractories,   \begin{equation}
    1+e\cos\nu_\mathrm{max}=\frac{a(1-e^2)}{r_\mathrm{max}}\qquad,
  \end{equation}
  where $a$ is the semi-major axis and $e$ the eccentricity. Using then Kepler's second law to obtain $\rd\nu/\rd t$, the integral becomes
  \begin{equation}
    \Delta s = \frac{2\alpha}{n\rho a^{16}}\frac{1}{(1-e^2)^{29/2}}
    \int_0^{\nu_\mathrm{max}}(1+e\cos\nu)^{15}\,\rd\nu\qquad,
\label{deltas}
  \end{equation}
  where $n$ is the mean motion of the orbit related to $a$ via Kepler's third law. The integral can be calculated in closed form as a polynomial expression involving the eccentricity and $\nu_\mathrm{max}$. We assume $z_0=10^7$\,kg\,s$^{-1}$ to be in agreement with the results of \citet{2003A&A...409..347K}, and integrated the radius loss (when relevant) over successive periastron passages for all particles involved in our simulations. This is a simplified model that deserves better treatment, but it is a convenient way to derive an estimate of the importance of evaporation from the computed dynamical evolution.  

Figure~\ref{ae_bpicbc_evap} shows the result of this computation. Both plots are actually identical to the 20\,Myr plots in Figs.~\ref{lac_truthc} and \ref{lac_rvc}, but dead particles (i.e. those that have been computed to be already fully evaporated at that time) are highlighted here in blue. In addition, we show the location of various MMRs with \bpc\ with red vertical bars. This shows how the inner disk is sculpted by those MMRs. We note three major facts. First, many particles close to the inner edge of the disk at 0.5\,au are already dead at 20\,Myr. This should not be surprising, as the average eccentricity they reach by secular perturbations by the two planets lets them constantly get below 0.4\,au and evaporate. Second, many particles that have been involved in the resonant eccentricity process are also dead. This also is not surprising, as they regularly enter the evaporation zone that eventually drives them to destruction. This is the real FEB phenomenon simulated here. Third, even if many resonant particles have already disappeared at $t=20\,$Myr, some of them are still present (black dots in the upper part of the plots). This shows that the FEB phenomenon is still active at $t=20\,$Myr.   

\begin{figure*}[tbp]
\makebox[\textwidth]{
\includegraphics[width=0.49\textwidth]{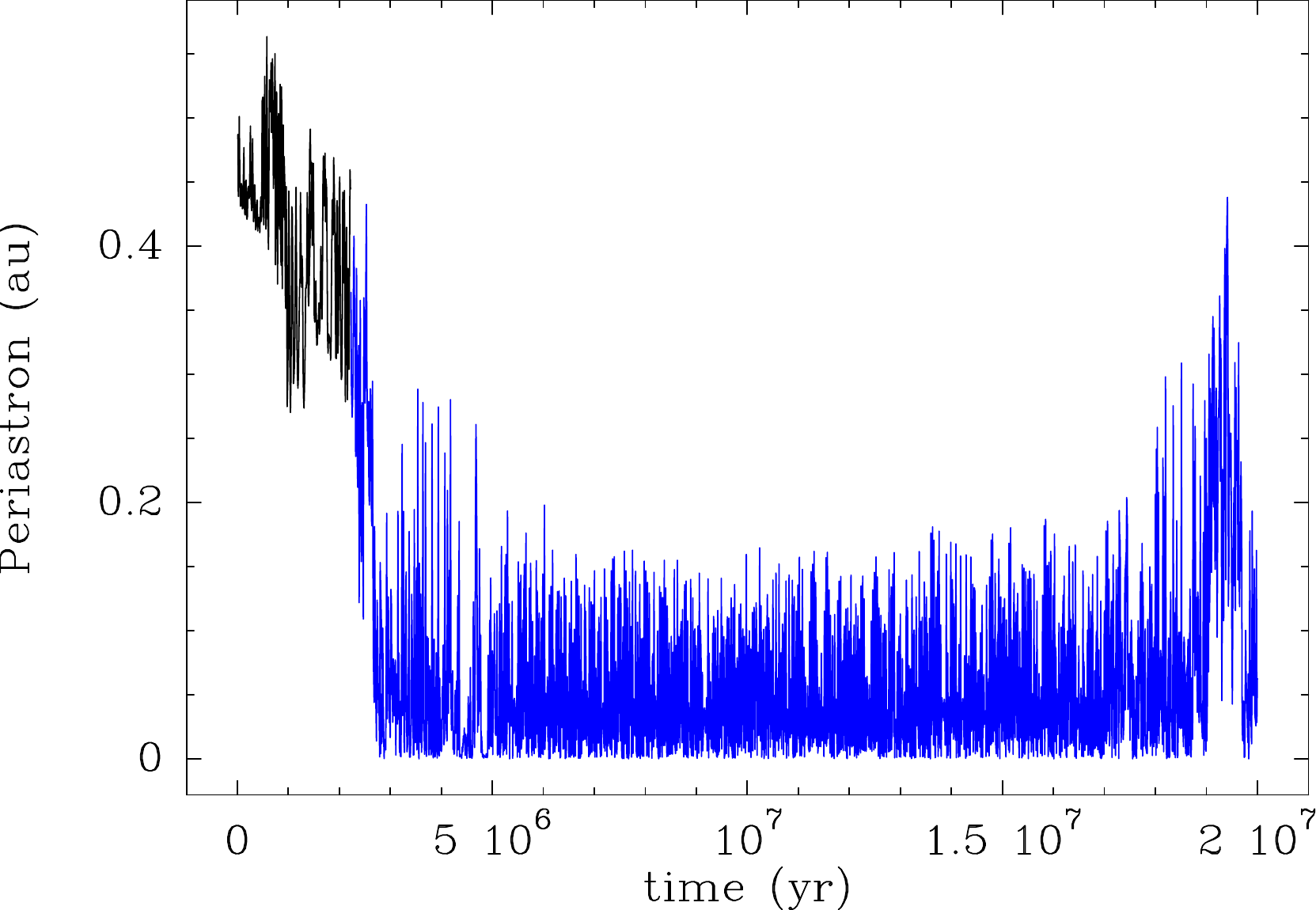} \hfil
\includegraphics[width=0.49\textwidth]{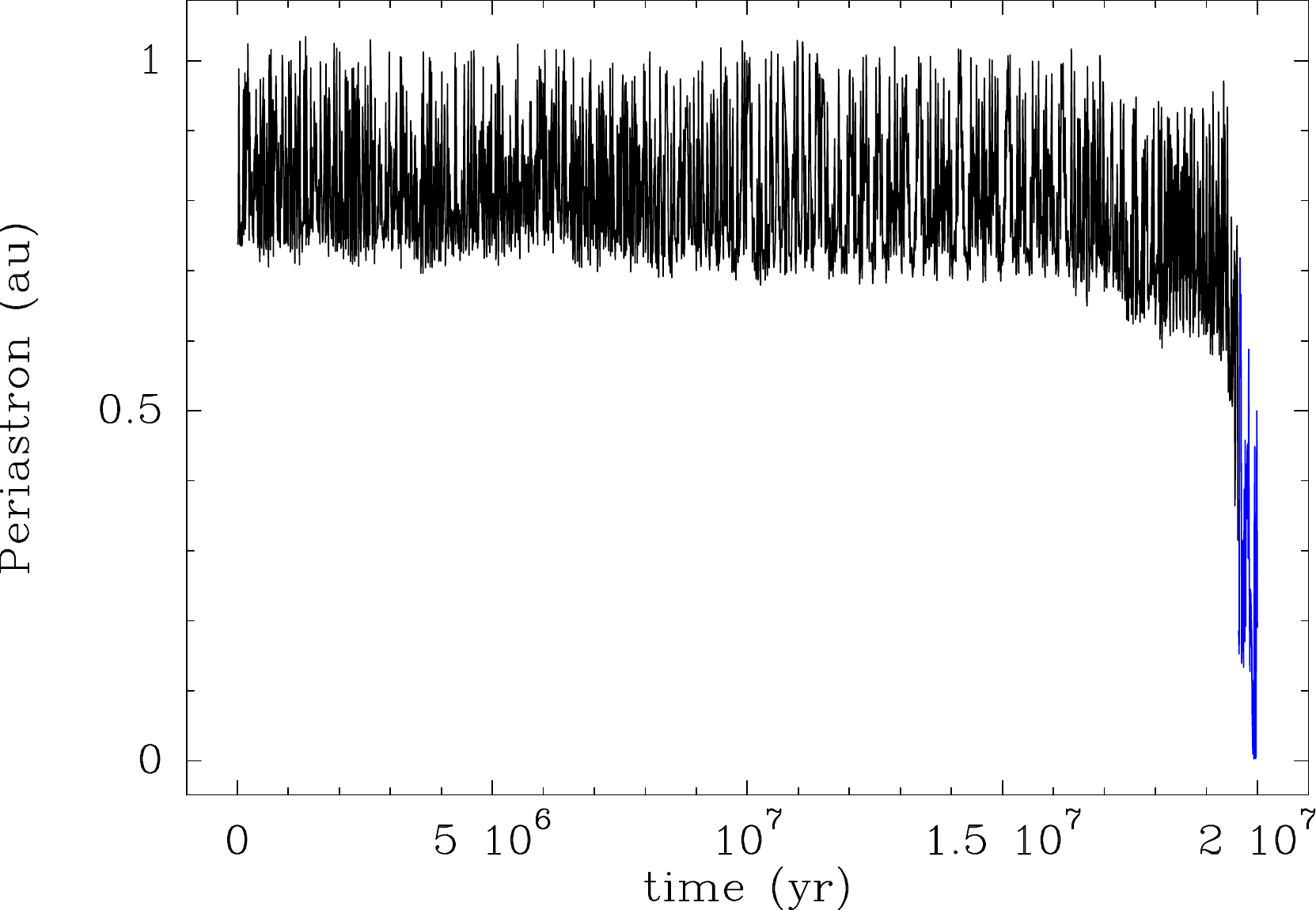}}
\caption[]{Temporal evolution of the periastron value of two individual bodies out of the simulations from Figs.~\ref{lac_truthc} and \ref{lac_rvc}. The colour convention follows that of Fig.~\ref{ae_bpicbc_evap}: the evolution appears in blue when the corresponding body is already evaporated.}
\label{peri_path}
\end{figure*}
The survival of bodies after $20\,$Myr could appear surprising as we could expect evaporation to act more quickly. This is actually not true close to the inner edge of the disk. A numerical application of Eq.~(\ref{deltas}) gives $\Delta s\simeq10^{-5}$--$10^{-4}$\,m at 0.4\,au. Therefore, assuming an orbital period of 0.27\,yr at $a=0.5\,$au, the whole radius loss after 20\,Myr ranges between 800\,m and 8\,km. This is enough to allow some of the bodies concerned to be fully evaporated, but not all, which is what Fig.~\ref{ae_bpicbc_evap} confirms. Evaporation is conversely much more efficient for resonant bodies when they reach high eccentricities ($\Delta s\simeq 1$\,m for a periastron of 0.2\,au). None of these bodies should be able to survive 20\,Myr. Hence, the presence of living bodies at high eccentricities in the MMRs at that time should be surprising.

The reason why some living bodies appear at high eccentricity even after 20\,Myr can be seen in Fig.~\ref{peri_path}, which displays the temporal evolution of the periastron of two particles out of the simulations. We adopt here the same convention as in Fig.~\ref{ae_bpicbc_evap} (i.e. the orbital path appears in blue when the bodies are fully evaporated). These parts of the curves should therefore be considered  virtual, as the corresponding bodies are already dead. Both particles are involved in the resonant process, but the decrease in periastron, a consequence of the eccentricity increase process, does not occur at the same time. The first particle (left plot) gets quickly involved in that process and disappears. The second particle (right plot) remains at low eccentricity (i.e. in the main disk) for almost all the simulation, and gets trapped in the same process only close to 20\,Myr and subsequently disappears. Hence fresh bodies permanently enter the FEB regime at any time and contribute to sustaining the FEB activity even even after 20\,Myr.

There are several reasons why some particles enter the FEB process after a long time. First, \citet{2000Icar..143..170B} showed (with the 4:1 resonance) that not all resonant particles are subject to the eccentricity increase process, but only those having a small enough resonant libration amplitude. Particles that have initially large libration amplitudes may stay at low eccentricity in the MMRs and never get involved in the FEB process, or possibly after some delay. Second, the location of the \bpc\ resonances themselves is subject to secular changes, thanks to mutual perturbations between the two planets. Consequently, particles that are initially not resonant may then be captured and enter the FEB regime, possibly after several million years.

We also note from Fig.~\ref{ae_bpicbc_evap} that some of the highest-order MMRs involved in the FEB mechanism seem to still be at the beginning of this process, even after 20\,Myr. This is for instance the case of the 13:2 and 16:3 MMRs, which seem not to have fully carved the corresponding zone in the disk and to be sending particles at high eccentricity that are still active. The reason is that the eccentricity increase process associated with those very high-order and weak MMRs is very slow compared to stronger ones, such as 7:1 and 8:1. Hence, after 20\,Myr, the associated resonant particles have still not evaporated.  

Finally, in addition to those various mechanisms, mutual scattering between planetesimals can also contribute to continuously replenishing the MMRs. This additional effect is not simulated here as our planetesimals are treated as massless particles. 
\section{FEBs velocity statistics}
\subsection{Theoretical background and semi-analytical study}
In \citet{2000Icar..143..170B} and \cite{2001A&A...376..621T} we   showed that the resonant mechanism supposed to be the source of FEBs towards \bp\ was able to trigger a non-axisymmetric infall of FEBs, thus generating  asymmetric statistics between blueshifted and redshifted spectral absorption events. In the single-planet model, this was fully constrained by the orientation of the planet's periastron. In comparison with the Doppler velocity statistics of observed events, which appear strongly biased towards redshifts \citep{1999MNRAS.304..733P}, it was even possible in the framework of that model to deduce a suitable value for the planet's longitude of periastron with respect to the line of sight that satisfactorily reproduced that statistics. This ability to trigger non-axisymmetric FEB velocity statistics was the main argument in favour of the resonant origin for FEBs in contrast with Kozai--Lidov resonance.

With the new model involving both planets, \bpb\ and \bpc, and FEBs arising from higher-order MMRs with \bpc, the picture is somewhat more complex, and it is worth wondering about the FEB velocity statistics it should generate (i.e. the radial velocities of all potential FEBs appearing in Figs.~\ref{lac_truthc} and \ref{lac_rvc} at the time they cross the line of sight). Following previous studies and \citet{2017A&A...605A..23P}, we first performed a semi-analytical analysis of the phenomenon. In \citet{1996Icar..120..358B}, portrait phases of the secular resonant Hamiltonian   indeed showed how resonant particles reach high eccentricities with constrained longitudes of periastron. We then tried to build similar maps adapted to the present model, and \begin{figure*}
\makebox[\textwidth]{
\includegraphics[width=0.49\textwidth]{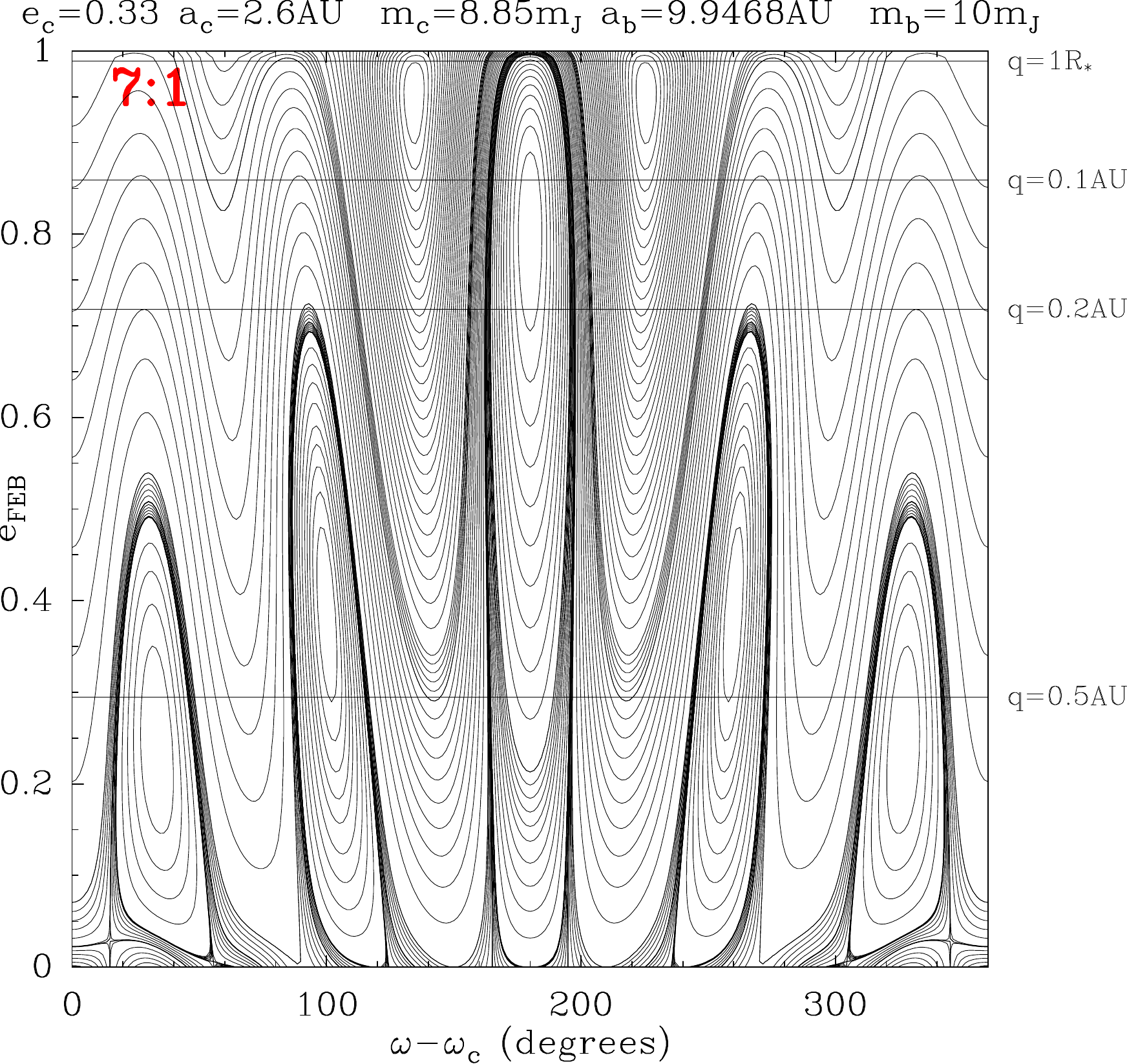} \hfil
\includegraphics[width=0.49\textwidth]{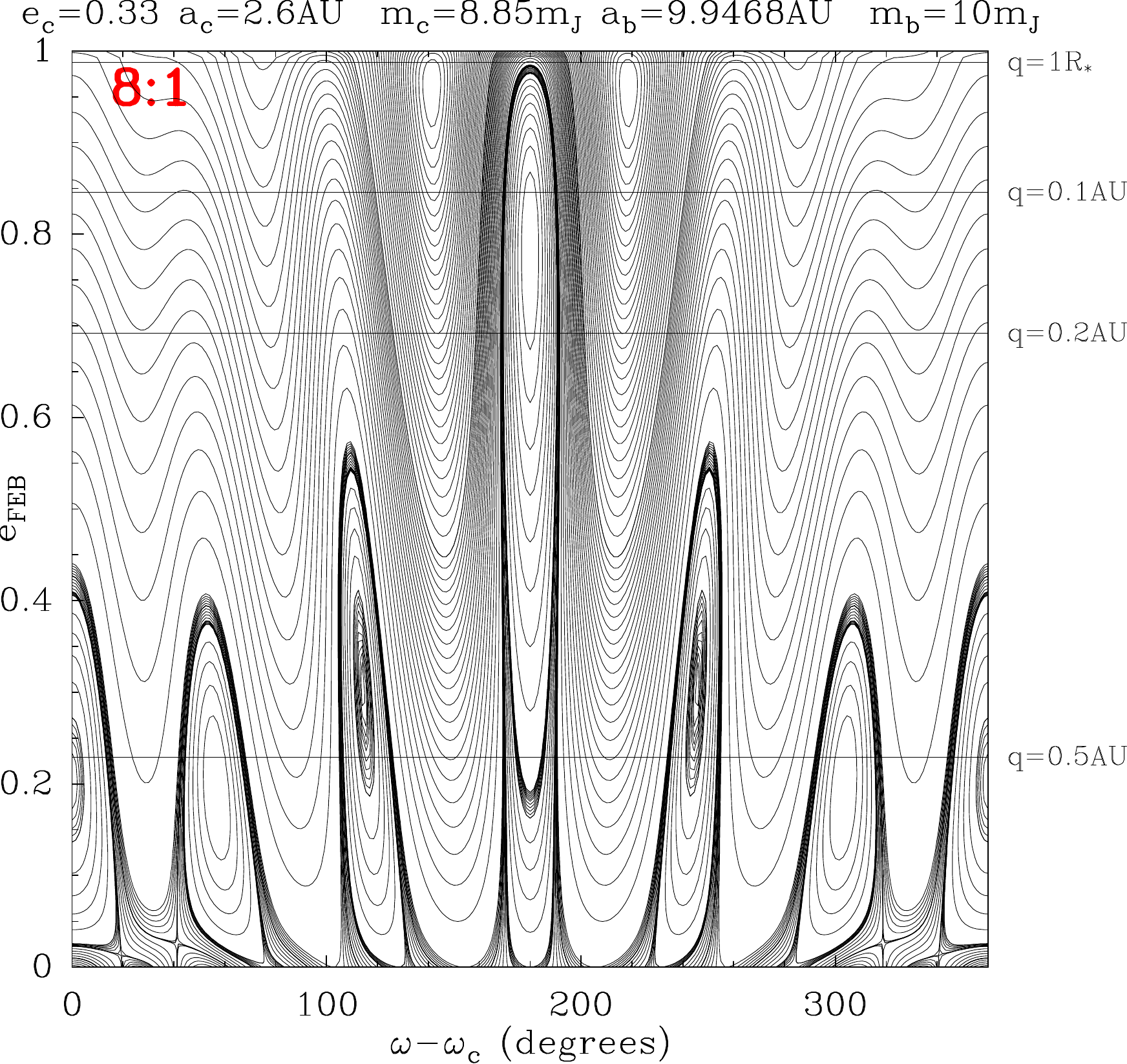}}
\caption[]{Phase portraits of the full secular Hamiltonian (\ref{hsec}) in $(\varpi-\varpi_c,e)$ space for particles trapped in 7:1 (left) and 8:1 (right) MMRs with \bpc\ having negligible resonant libration amplitude, taking into account perturbations by \bpb\ and General Relativity (see text). The horizontal bars denote the location of orbits with specific periastron values, down to the stellar surface ($1\,R_*$).}
\label{hsec_map}
\end{figure*}
considered a massless particle orbiting \bp, perturbed by both planets, \bpb\ and \bpc. The instantaneous interaction Hamiltonian reads
\begin{equation}
  H=-\frac{GM}{2a}+H_b+H_c\qquad,
\end{equation}
with
\begin{eqnarray}
  H_c & = & -Gm_c\left(\frac{1}{|\vec{r}-\vec{r_c}|}-\frac{\vec{r}\cdot
    \vec{r_c}}{r_c^3}\right)\quad\mbox{and}\quad\nonumber\\
  H_b & = & -Gm_b\left(\frac{1}{|\vec{r}-\vec{r_b}|}-\frac{\vec{r}\cdot\vec{r_b}}{r_b^3}\right)\;,
\end{eqnarray}
where $M$ is the mass of the star; $m_c$ and $m_b$ are the masses of the two planets; and $\vec{r}$, $\vec{r_c}$, and $\vec{r_b}$ are respectively the heliocentric radius vectors of the particles and the planets. The secular Hamiltonian is obtained taking the time average of this Hamiltonian over the various orbital motions. Let us consider that the particle is trapped in \mbox{($p+q:p$)} MMR with \bpc\ and not in any MMR with \bpb\ (unless a very high-order weak one, which we  neglect). Hence the interaction Hamiltonian with \bpb\ $H_b$ can be averaged over the orbital motions of the particle and \bpb\ independently. This reads
\begin{equation}
H_b=-\frac{Gm_b}{4\pi^2}\oint\!\!\oint
\left(\frac{1}{|\vec{r}-\vec{r_b}|}-\frac{\vec{r}\cdot
\vec{r_b}}{r_b^3}\right)\,\rd l\,\rd l_b\;,
\end{equation}
where $l$ and $l_b$ are the mean
anomalies of the particle and \bpb, respectively. To the lowest (quadrupolar) order in semi-major axis $a/a_b$ ratio (we implicitly assume $a\ll a_b$), and assuming full coplanarity, this reads in closed form
\begin{equation}
  H_{\mathrm{sec},b,0}=-\frac{Gm_b}{4}\frac{a^2}{a_b^3}\frac{1+(3/2)e^2}{(1-e_b^2)^{3/2}}\qquad,
\end{equation}
where $e$ and $e_b$ are the eccentricities of the two bodies. As explained in \citet{2016A&A...590L...2B} and \citet{2016AJ....151...22B}, this expression is only valid in the case of a fixed orbit for \bpb. Considering \bpb's orbital precession under perturbations by \bpc, the full expression is
\begin{equation}
  H_{\mathrm{sec},b}=H_{\mathrm{sec,b,0}}+\sqrt{aGM}\left(\sqrt{1-e^2}-1\right)\dot{\varpi_b}\qquad,
\end{equation}
where $\dot{\varpi_b}$ is the periastron precession velocity of \bpb.

A similar treatment must be done with the interaction $H_c$ with \bpc, but an independent averaging over the orbital motions of the particle and of \bpc\ cannot be done because of the MMR configuration. A canonical transformation must first be done involving the critical argument of the resonance
\begin{equation}
\sigma=\frac{p+q}{p}\,\lambda_c-\frac{p}{q}\,\lambda-\varpi\;.
\end{equation}
Here again, the precession of \bpc's longitude of periastron introduces additional terms in the canonical transformation \citep[see][for details]{2016A&A...590L...2B}. The Hamiltonian then becomes
\begin{equation}
H_\mathrm{res} = H_c+\dot{\varpi_c}\sqrt{aGM}
\left(\frac{p+q}{p}-\sqrt{1-e^2}\right)-\frac{p+q}{p}\,n_c\sqrt{aGM}\:,
\end{equation}
with $n_c$ being \bpc's mean motion. The final averaging of $H_c$ is then done over $\lambda_c$ only at constant $\sigma$. As we are considering orbits with rather small semi-major axes, we must also add, following \citet{2017A&A...605A..23P}, the first order post-Newtonian correction due to General Relativity:
\begin{equation}
  H_\mathrm{GR}=\frac{1}{c^2}\left(\frac{GM}{a}\right)^2\left(\frac{15}{8}-\frac{3}{\sqrt{1-e^2}}\right)\qquad.
\end{equation}
It should be noted that that expression is already averaged over the particle's orbital motion. Following again \citet{2017A&A...605A..23P}, we  neglect the extra contribution due to the rotational bulge of the star. The resulting secular Hamiltonian is therefore
\begin{equation}
  H_\mathrm{sec} = -\frac{GM}{2a}+H_{\mathrm{sec},b}+H_\mathrm{GR}+\oint H_\mathrm{res}\,\rd\lambda_c\qquad.
\label{hsec}
\end{equation}
Now, to first order in perturbations between \bpc\ and \bpb, we derive
\begin{eqnarray}
  \dot{\varpi_b} & = &\frac{3n_b}{8}\frac{a_c^2}{a_b^2}\frac{m_c}{M+m_b}\frac{3e_c^2+2}{(1-e_b^2)^2}\qquad,\\
  \dot{\varpi_c} & = & \frac{3n_c}{4}\frac{a_c^3}{a_b^3}\frac{m_b}{M+m_b}\frac{\sqrt{1-e_c^2}}{(1-e_b^2)^{3/2}}\qquad.
\end{eqnarray}
Assuming coplanarity, the first three terms of $H_\mathrm{sec}$ alone would result in a secularly constant semi-major axis and constitute a one degree of freedom Hamiltonian.  However the presence of the fourth term does not easily reduce to one degree of freedom, as the averaging process is done for constant $\sigma$ and not constant $a$. As detailed by \citet{1993Icar..102..316M} and \citet{2017A&A...605A..23P}, the secular evolution is characterized by another adiabatic invariant $J$ corresponding roughly to the amplitude of the $\sigma$-libration. As pointed out by \citet{1996Icar..120..358B} and \citet{2017A&A...605A..23P}, the study is considerably simplified if we consider orbits with negligible $\sigma$-libration amplitude. In this case, the semi-major axis remains secularly constant, and the Hamiltonian reduces to one degree of freedom. Portrait phases in $(\varpi-\varpi_c,e)$ space can be drawn, as was done in \citet{2016A&A...590L...2B}, for instance.

Figure~\ref{hsec_map} shows two such portrait phases for particles trapped in 7:1 and 8:1 MMRs with \bpc, as they are the main contributors of FEBs in our simulations. The maps were made assuming solution \#2 of Table~\ref{orbits} (non-resonant), but assuming the other solution does not reveal significant changes. Basically, these maps appear more complex than those corresponding to the 4:1 and 3:1 MMRs presented in \citet{1996Icar..120..358B}. This is due to the high order or the MMRs under consideration here, and to the significant eccentricity of \bpc. We note that the maps were computed numerically with no specific assumption about that eccentricity. They may actually be compared to similar maps in \citet{2016A&A...590L...2B}. Nonetheless, particles starting at low eccentricity have many possible routes to reach high eccentricities and become FEBs following the level curves drawn here. Hence, the quoted MMRs are plausible theoretical FEB generators. Moreover, the eccentricity increase occurs near specific values of $\varpi-\varpi_c$, suggesting a non-axisymmetric infall of FEBs as depicted above. However, the relative complexity of the phase portraits shown here, combined with the fact that many MMRs may be simultaneous FEB contributors, and taking into account the intrinsic chaotic nature of the real resonant motion suggests that we should return to the numerical study.
\subsection{Numerical exploration}
\begin{figure*}
\makebox[\textwidth]{
\includegraphics[width=0.49\textwidth]{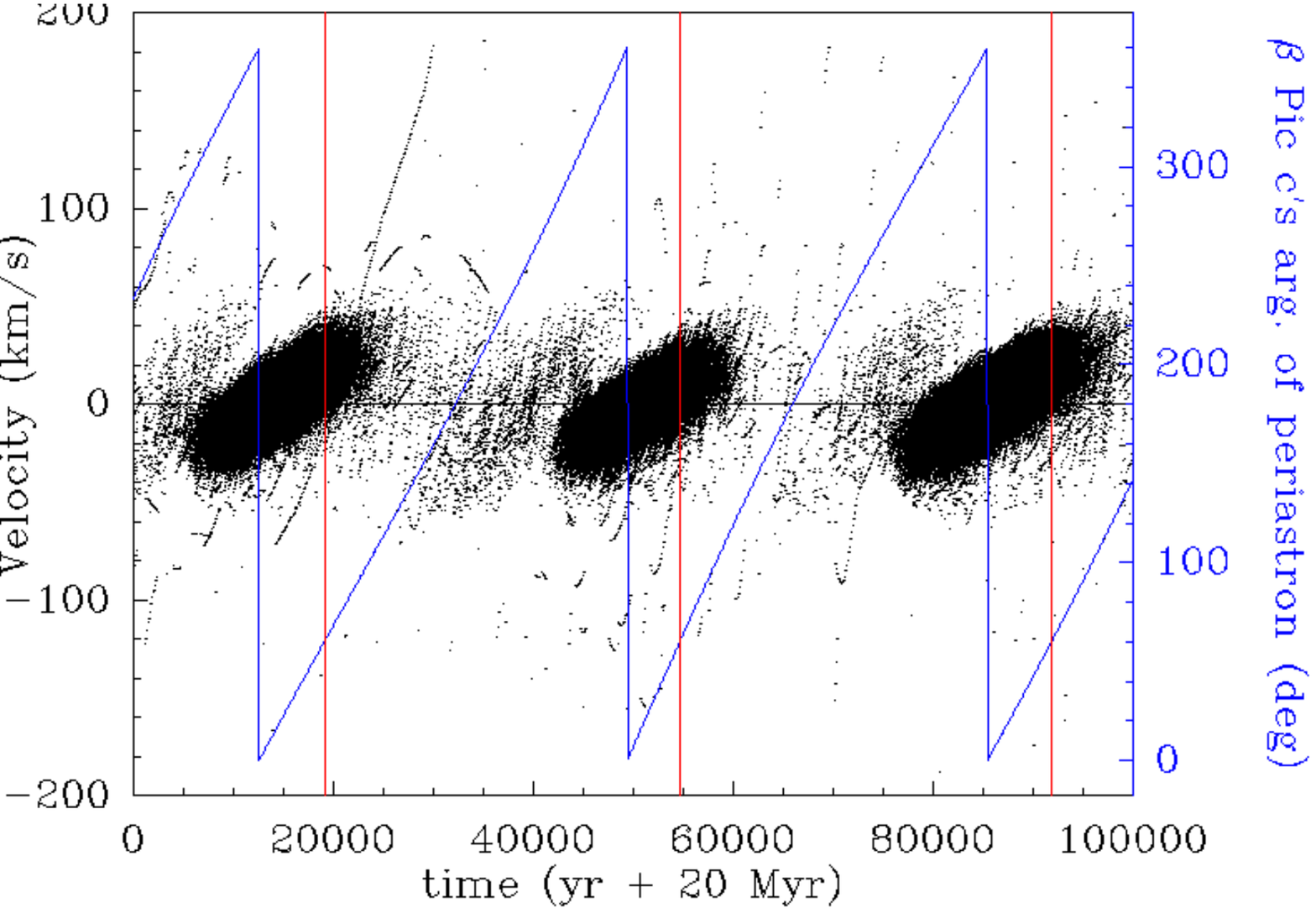} \hfil
\includegraphics[width=0.49\textwidth]{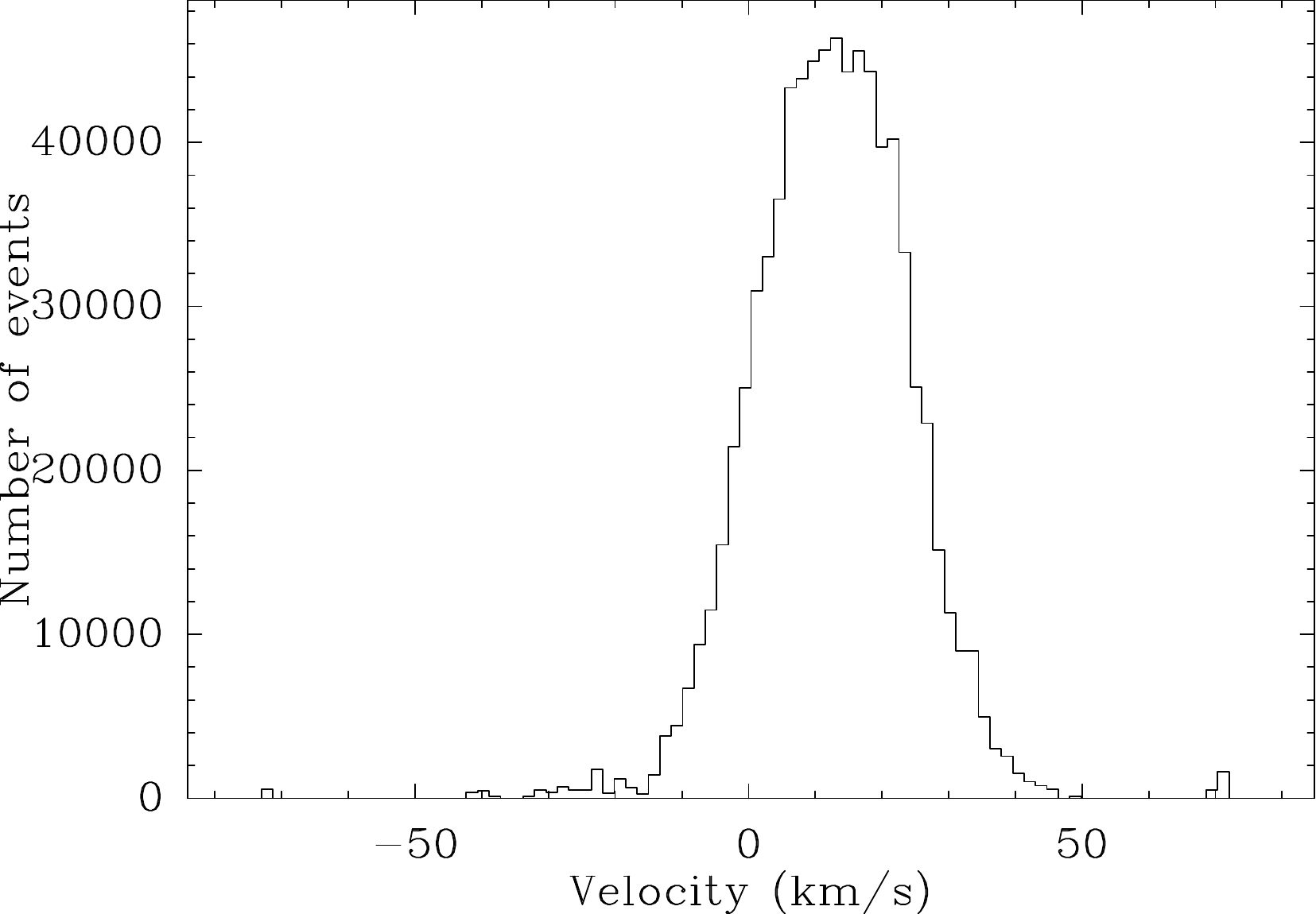}}

\caption[]{Statistical description of the simulated FEB spectral events generated from the simulation described in Fig.~\ref{lac_truthc}, and corresponding to solution \#2 from Table~\ref{orbits}. \textbf{Left:} Doppler velocities of the FEB events occurring in the first $10^5\,$yr after the end of the simulation. The velocity scale is on the left side of the plot; positive velocities correspond to redshifts. The simultaneous evolution of \bpc's argument of periastron $\omega_{c,\mathrm{sky}}$ (see text) is superimposed in blue, with the corresponding scale in blue on the right side of the plot. The red bars highlight epochs where $\omega_{c,\mathrm{sky}}$ matches the fitted value from Table~\ref{orbits}. \textbf{Right:} Histogram of FEB velocities occurring $\pm 100\,$yr around epochs corresponding to the red bars.} 
\label{modul_truthc}
\end{figure*}
\begin{figure*}
\makebox[\textwidth]{
\includegraphics[width=0.49\textwidth]{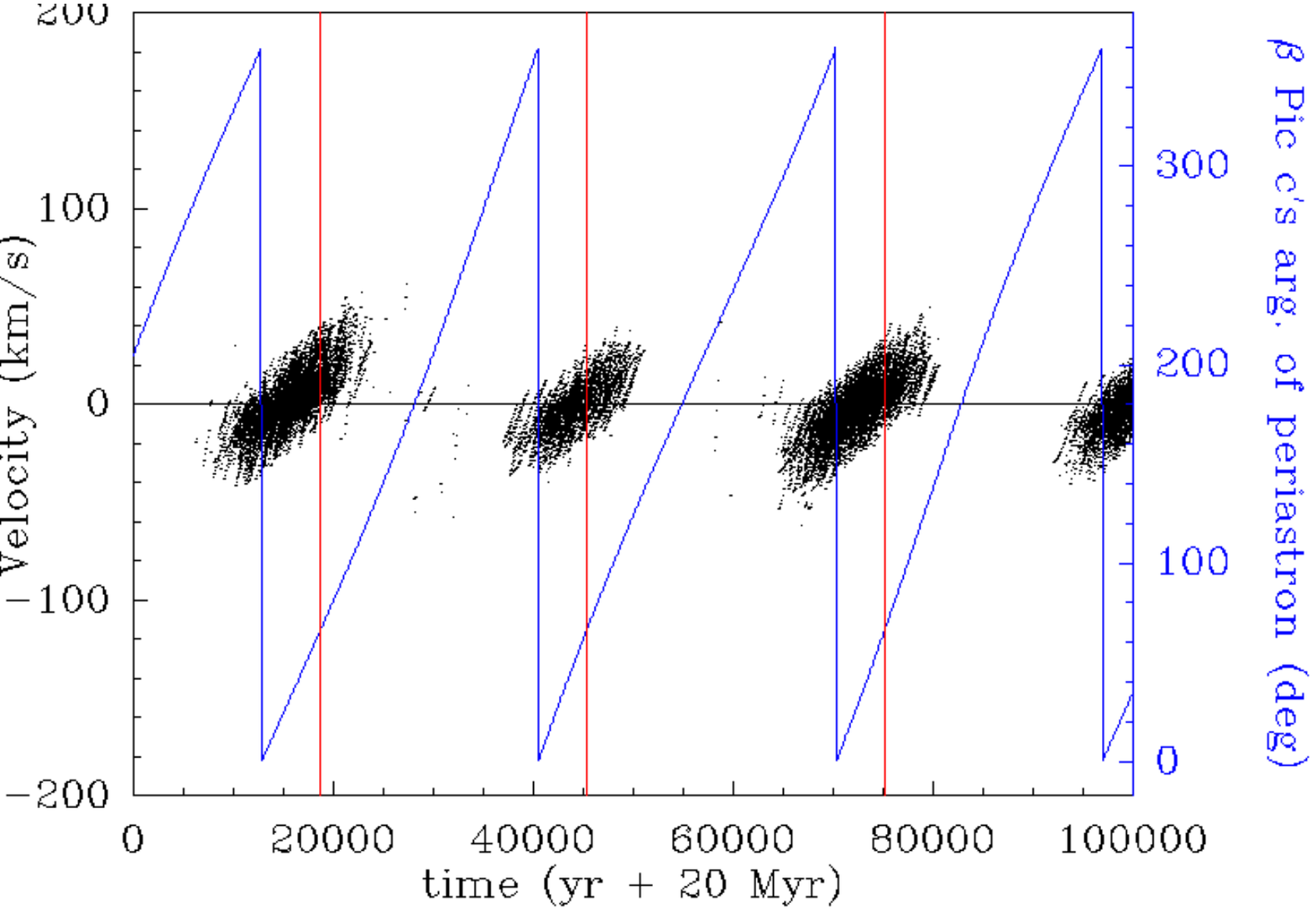} \hfil
\includegraphics[width=0.49\textwidth]{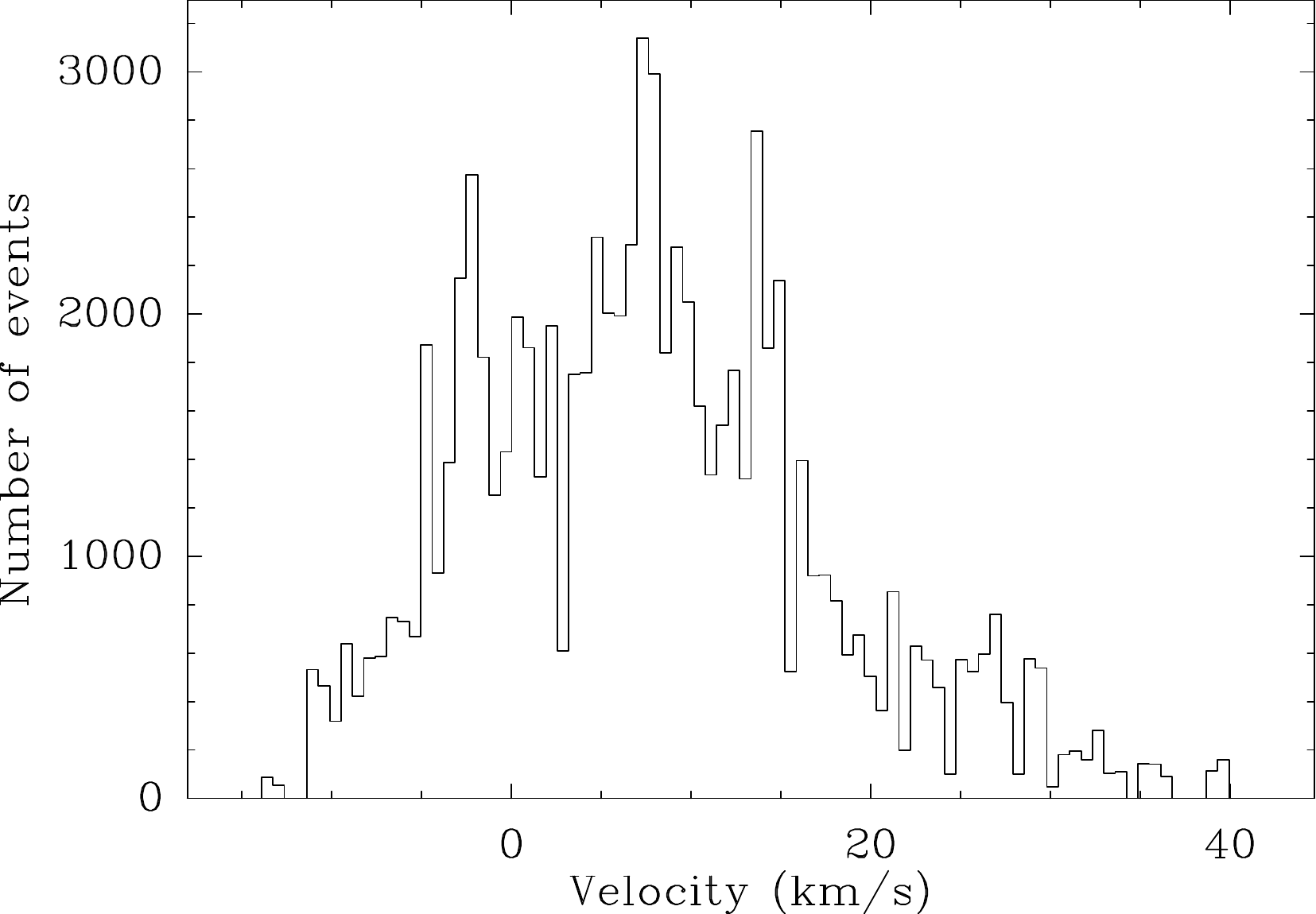}}
\caption[]{Same as Fig.~\ref{modul_truthc}, but for the simulation described in Fig.~\ref{lac_rvc}, and corresponding to solution \#1 from Table~\ref{orbits}.}  
\label{modul_rvc}
\end{figure*}
In this section we return to the simulations of the inner ring of particles depicted in Sect.~\ref{inner}. Figures~\ref{lac_truthc} and \ref{lac_rvc} show that high-order MMRs with \bpc\ may constitute active sources of FEBs, and the above semi-analytical study gives it a theoretical background, suggesting in addition that the FEB infall may be non-axisymmetric. We  therefore  investigate the statistics of FEB Doppler velocities that should be expected from those simulations; however, this is  not as straightforward as it could first seem. First, the simulations were run in the invariable plane of the two-planet system, and did not assume any specific direction for the line of sight. Second, as the simulations extended over 20\,Myr, the orbital elements of all particles were stored in the output every 10,000\,yr to save disk space. This sampling was sufficient to reveal the reality of the mechanism, as can be seen from Figs.~\ref{lac_truthc} and \ref{lac_rvc}, but   not to build  the  relevant statistics of all periastron passages of all particles. Hence, we decided to let the simulations run 200,000 yr after their end at 20\,Myr, but we began storing  orbital elements every 100\,yr and interpolating all periastron passages occurring in between. 

Next, we needed to fix a direction for the line of sight to be able to compute FEB velocities at the time they cross it. We first decided to assume for simplicity that the line of sight lies in the orbital plane of \bpc\ (i.e. assimilating its orbital inclination to $90\degr$ with respect to the sky plane). As can be seen from Table~\ref{orbits} and \citet{2021A&A...654L...2L}, the actual value is close to $89\degr$, so that the error on the projected velocities should not be significant. Then we decided to fix the line of sight so as to coincide with the $OX$-axis of our referential frame. We note that this arbitrary choice does not mean assuming anything on the orbital configuration of \bpc\ relative to the line of sight, as its orbit is expect to precess under the perturbations of \bpb. In this context, and remembering that the orbital conventions of \citet{2021A&A...654L...2L} and \citet{2020AJ....159...89B} assume an $OZ$-axis pointing away from the line of sight, it is straightforward to derive the following relation between \bpc's longitude of periastron $\varpi_c$, as computed from our simulations, and its argument or periastron $\omega_{c,\mathrm{sky}}$ following the quoted conventions:
\begin{equation}
  \omega_{c,\mathrm{sky}}=\varpi_c-\frac{\pi}{2}\qquad.
\end{equation}
\citet{2020A&A...642A..18L} assume an opposite convention with an $OZ$-axis pointing towards the observer. This results in a $180\degr$ shift in the value of the argument of periastron $\omega$, as can be seen by comparing the orbital fits between \citet{2020A&A...642A..18L} and \citet{2021A&A...654L...2L}.    

Figures~\ref{modul_truthc} and \ref{modul_rvc} show the result of this computation for the simulations described in Figs.~\ref{lac_truthc} and \ref{lac_rvc}, corresponding respectively to solution \#2 (non-resonant) and \#1 (partly resonant) from Table~\ref{orbits}. In each case the expected velocities of all FEB events simulated are plotted as a function of time in the first $10^5\,$yr of the extended simulation. In this computation, however, we only took the still-not-fully-evaporated planetesimals into account (black dots in Fig.~\ref{ae_bpicbc_evap}). We also kept computing the evaporation of planetesimals during the extended simulation.  

The velocity statistics at any epoch can therefore be viewed directly from the plots. A clear temporal modulation over a few $10^4\,$yr appears, which exactly corresponds the precession of \bpc's periastron. To highlight this, the temporal evolution of $\omega_{c,\mathrm{sky}}$ is superimposed in blue to the plots. Then, red bars corresponding to epochs when $\omega_{c,\mathrm{sky}}$ matches the fitted values from Table~\ref{orbits} are overplotted. Finally, the right plots of both figures display the extracted histograms of  Doppler velocities of FEB events occurring $\pm100\,$yr around epochs, corresponding to the red bars in the left plots. These histograms simulate the expected velocity statistics of FEBs events for the present-day configuration of \bpc\ with respect to the line of sight.

The first thing we note is a global agreement with the observed statistics of FEB velocities (i.e. a predominance of redshifts, although blueshifts are present). The histograms in Figs.~\ref{modul_truthc} and \ref{modul_rvc} actually  compare very well with the observational statistics of \citet{2014Natur.514..462K} and \citet{2019MNRAS.489..574T}. The only noticeable difference is perhaps the absence of simulated high-velocity events ($\ga\pm 100\,$km\,s$^{-1}$) compared to the observations. We point out here that the number of such events is probably highly underestimated in our simulation. Higher-velocity events correspond to FEBs crossing the line of sight very close to the star ($\la0.1\,$au). Our evaporation rules (Sect.~\ref{evap}) are based on the simulations of \citet{2003A&A...409..347K} for average silicate material. However,  the same work showed that, depending on the chemical composition of the body, the sublimation occurs at different distances. In particular, FEBs made of graphite should be able to resist as close as 0.1\,au. \citet{2006ApJ...643..509F} and \citet{2011ApJ...729..122B} indeed showed that carbonaceous FEBs should be more able to retain their evaporated metallic compounds and generate observable signatures in absorption than others.

These trends are clear in both solutions, but the number of FEB events is obviously larger in the non-resonant configuration. The reason is that in the resonant case more FEBs miss the line of sight, due to larger chaos in their inclination evolution. \citet{2007A&A...466..201B} actually showed that the resonant FEB evolution is characterized by large inclination oscillations in the high-eccentricity phase. Here the fact that the FEBs are sometimes locked in MMR configuration with both planets causes larger chaos in that evolution, resulting in fewer events.

The statistics of FEB velocities is expected to change with time as \bpc's periastron precesses. The present-day configuration favours redshifts, but blueshifts should be dominant at other epochs. There are also epochs where FEBs events are much less numerous. This corresponds to configurations where the periastron of \bpc\ is oriented in such a way that most FEB periastron passages occur behind the star as seen from the Earth. 

It should also be noted here that all the cycles over $10^5\,$ in Figs.~\ref{modul_truthc} and \ref{modul_rvc} are not identical. This highlights the discrete role of \bpb. Although the periodic configuration of \bpc\ relative to the line of sight dominates the process, the configuration relative to \bpb\ does not evolve with the same periodicity, resulting in small changes in the successive cycles. 
\section{Discussion}
The presence of both \bpb\ and \bpc\ orbiting \bp\ drives us to a totally renewed view of the FEB generation mechanism. All previous studies pointed out the role of a distant planet that closely matches \bpb\ as being responsible for the whole process via its low-order inner MMRs. The unexpected presence of \bpc\ changes this picture. MMRs with \bpb\ can no longer be sources of FEBs as the corresponding regions are dynamically unstable. Conversely, much higher-order MMRs between $\sim0.6\,$au and $\sim 1.5\,$au with \bpc\ now appear   to be good candidates.

As in the previous model involving one perturbing planet, the FEB infall towards the star is not axisymmetric. The result is asymmetric statistics of FEB velocities at the time they cross the line of sight, favouring redshifs or blueshifts depending on the configuration of \bpc's periastron with respect to the line of sight. Our simulations show that the present-day fitted configuration actually favours redshifts, in agreement with observations. The simulated histograms in Figs.~\ref{modul_truthc} and Fig.~\ref{modul_rvc} confirm  this general trend. Several MMRs actually contribute to the FEB population, and many dynamical routes can lead to a  FEB state inside individual MMRs (see Fig.~\ref{hsec_map}). This could provide a dynamical origin for the different families of FEB events quoted by \citet{2014Natur.514..462K}, thanks to a careful analysis of the events statistics.

The fact that planetesimals may spend a significant time in the disk before being captured into a MMR and involved in the FEB generation mechanism explains the duration of the mechanism. \citet{2001A&A...376..621T} showed  in the previous model with only \bpb, that the 4:1 and 3:1 MMRs with \bpb\ quite rapidly cleared out, causing the FEB process to nearly stop after 1--2 Myr. Collisions among planetesimals next to those MMRs had been invoked as a way to replenish the MMRs and sustain the process. Here thanks to its distant perturbations added to the global action of \bpc, \bpb\ helps the resonant FEB generation mechanism to   still be active at the present age of \bp. This activity is  not expected to last for ever, however. At some point the whole planetesimal disk further out than $\sim 0.6\,$au will be cleared out, and the FEB activity will cease. But the point here is that this is expected to happen later than the present age of \bp. Letting the simulation run (far) beyond 20\,Myr should help in specifying this issue. The perturbing action of \bpb\ also influences the observability of the FEBs in absorption. \bpb\ causes \bpc's periastron to secularly precess, and subsequently the statistics of FEB velocities to periodically change. The present-day configuration of \bpc's periastron with respect to the line of sight is compatible with redshifts, but this is about to evolve with time. 

The new model also questions the physical nature of the FEBs themselves. In the preceding picture, FEBs were supposed to originate from regions located between $\sim 3.5\,$au and $\sim 4.5\,$au (4:1 and 3:1 MMRs with \bpb). Depending on the exact location of the ice line in the \bp\ system, those FEB were likely to be at least partly icy. Ices were  invoked in previous studies \citep{1990A&A...236..202B,2003A&A...409..347K} as a necessary source of volatile material in the FEB comas that prevents the metallic ions from being immediately blown away by the stellar radiation pressure, letting them first expand radially around the nucleus. This is required because it allows the ion clouds to cover a significant part of the stellar surface, and subsequently renders the FEBs spectral components visible. Here our revised model puts the FEB reservoir between $\sim 0.6$\,au and $\sim 1.5\,$au (i.e. much closer to the star). At that distance, the planetesimals are not expected to be icy at all. Ices may  not be necessary, however. In early studies \citep{1990A&A...236..202B,1996A&A...310..181B}, volatile compounds were introduced as a braking agent for metallic ions as they are not affected by any noticeable radiation pressure from the star, contrary to metallic ions. Collisional activity between various species is sufficient to render the gas self-braking where the density is high enough (i.e. close to the nucleus). \citet{2006ApJ...643..509F} showed in a detailed analysis that if the gas surrounding the FEB nucleus (arising from dust grain sublimation) contains enough carbonaceous compounds, the gas can be self-braking in the same way as before, but with no volatiles. The presence of solid material orbiting inside \bpc's orbit is also supported by recent independent observations. In a careful analysis of \bp's infrared spectrum and its associated silicate features, \citet{2022ApJ...933...54L} note the presence of a weak excess at 5$\mu$m tentatively attributed to a hot dust population at $\sim 600\,$K, located within 0.7\,au from the star. This hot dust could obviously be related to our FEB reservoir. Similarly, the stable gas component observed at rest with respect to the star could also originate from this reservoir. As shown by \citet{1998A&A...330.1091L} in an early study, this component is likely to arise from the FEBs themselves as the corresponding gas amount needs to be continuously replenished.

Another open issue is the survival of this planetesimal population to collisional activity. This is particularly relevant here as the  mean eccentricities generated by the planetary perturbations on this population are large (see Fig. \ref{ae_bpicbc_evap}). Actually, the main question is not whether the  population   survives collisions, but how long the population should survive. The only requirement is that it should be able to survive until the current age of the star. Investigating this issue is beyond the scope of the present paper, and will require a dedicated study. There are nevertheless a few clues indicating that its survival could be possible. First, as noted above, there is observational evidence for an existing dust population inner to 1\,au \citep{2022ApJ...933...54L}; second, the total population required to generate the observed FEB rate might not be that high as i) given the large number of MMRs that are active source of FEBs, up to $\sim10$\%\ of the total population may be affected by the resonant process and  ii) thanks to the eccentricities reached (up to $\sim0.3$), even non-resonant FEBs can participate to some extent in the FEB activity; and  third, at the same time the FEB progenitor population reaches a significant eccentricity dispersion, a similar spread happens to the inclinations. This involves in particular resonant FEBs, which can reach inclinations up to several tens of degrees. This behaviour was already noted in \citet{2007A&A...466..201B}, and can be explained as a combination of Kozai resonance and MMRs. Nonetheless, if a given planetesimal population is able to reach a large inclination dispersion, its collisional activity should be reduced compared to the same population confined to a flattened disk. As a final argument, it can be seen that collisions among planetesimals also naturally contribute to replenishing the MMRs from regions with adjacent semi-major axes, and thus sustain the FEB activity.

Finally, the last open issue is the presence of additional planets in the \bp\ system that could potentially affect the secular dynamics of the already-known planets, and subsequently the dynamics of  FEBs. Our feeling is that the present model should be fairly robust to the presence of other planets. The simulations presented here show that there is not much space for additional planets closer to $\sim 20\,$au as these would presumably be dynamically unstable. Planets located further out could conversely be present in the \bp\ disk. \citet{2020A&A...642A..18L} indeed give observational upper limits showing that planets up to $\sim 1\,\mjup$ in the 20--40\,au region could still remain undetected. However,  an additional planet, presumably less massive than \bpb\ and \bpc, and located much further out, is likely to have very little impact on the dynamics of planetesimals located around 1\,au. Even \bpc's periastron precession velocity should be only marginally affected. Hence, the presence of additional planets in the \bp\ system is not expected to significantly affect the present model.
\section{Conclusions}
We presented a renewed model for the dynamical origin of the FEBs in the \bp\ system. In the presence of both \bpb\ and \bpc, MMRs with \bpb\ located at 3--5\,au can no longer be a valuable source of FEBs as this region is intrinsically dynamically unstable. Higher-order MMRs with \bpc\ located much closer to the star constitute conversely a new potential source, their efficiency being furthermore enhanced by the more distant action of \bpb. In this context, a significant portion of the planetesimal disk between 0.6\,au and 1.5\,au is potentially capable of becoming FEBs, in contrast to the previous model where   only bodies initially trapped in one or two MMRs under consideration were able to do so. This causes the FEB activity in the \bp\ system to potentially last much longer than initially estimated, and to remain observable today at \bp's present age. Moreover, the present-day configuration of the two orbits relative to the line of sight is able to reproduce the observed statistics of FEB velocities that shows a clear trend favouring redshifts, and thus reinforcing the plausibility of the scenario. The main parameter controlling this process is actually the orientation of \bpc's periastron relative to the line of sight, and the current fitted orientation is compatible with a predominance of redshifts.

Many pending issues are still present in the new model and will require more dedicated studies. The statistics of FEB velocities needs to be more precisely simulated in the framework of the new model to more specifically address the issue of differentiated FEB families. This will be the purpose of future modelling work involving simulations with many more particles. Moreover, FEB progenitors arising from regions around ~1\,au are presumably not icy. As quoted above this is not a huge difficulty as carbon is likely to play a  braking role similar to that played by volatiles (previously invoked) for metallic ions in FEB comas. However, the exact physics involved here needs  to be modelled in the same way as was done with volatiles in early studies \citep{1990A&A...236..202B,1996A&A...310..181B}.

Finally, an observational test could be proposed. If the FEBs are actually not icy, it should be possible in principle to distinguish them from icy FEBs via specific spectroscopic studies. Interferometric observations have the capability to resolve the inner regions at sub-au scales where the dust generated by the FEB lies \citep{Defrere2012} to shed light on  its composition and water content, for instance using the low- to medium-resolution spectroscopic capability at L- and M-band (3 to 5\micron) of the VLTI/MATISSE instrument \cite[e.g.][]{Kokoulina2021}. 
%
\begin{acknowledgements}
All (or most of) the computations presented in this paper were performed using the GRICAD infrastructure (https://gricad.univ-grenoble-alpes.fr), which is supported by Grenoble research communities.    

This work used astrometric measurements collected at the European Southern Observatory under ESO large programme ExoGRAVITY (ID 1104.C-0651).

This project has received funding from the European Research Council (ERC) under the European Union's Horizon 2020 research and innovation programme (COBREX; grant agreement n$^\circ$885593).

S. L. acknowledges the support of the French Agence Nationale de la Recherche (ANR), under grant ANR-21-CE31-0017 (project ExoVLTI).
\end{acknowledgements}
%
\bibliographystyle{aa}
\bibliography{bpbiblio}
\end{document}